\documentstyle[11pt]{article}

\begin{document}

\pagestyle{myheadings}
\markboth{Draft  }{\today $\;\;\;$  { \sf }}

\begin{center}
\LARGE
Quark Cluster Model Equations for $\beta\beta$ Decay \end{center}
\begin{center}

%\vskip 2truecm
\large
\vskip .6truecm
D. Strottman$^a$ and J. D. Vergados$^b$\\
%Theoretical Division\\
%Los Alamos National Laboratory, Los Alamos, NM 87545

$^a$ Theoretical Division, T-DO, LANL, P.O. Box 1663,
Los Alamos, N.M. 87545.\\
$^b$Physics Department, University of Ioannina, Gr 451
10, Ioannina, Greece.

\normalsize
\vskip 0.5truecm
\centerline{  \today} 
\vskip 0.5truecm
{\bf Abstract}
\end{center}
In a separate paper we have discussed the possibility that six quark clusters can affect the rate of double-beta decay.  In this paper we present the notation and some of the formulae that are needed to carry through such calculations.
\tableofcontents
\listoftables
%\listoffigures

%\newpage

\section{Introdction}
Multi-quark systems are Fermions containing more than three quarks  and mesons containing more pairs than one quark one anti-quark.    Such multi-quarks  were later discovered experimentally, the penta-quark \cite{pentaq15,pentaq16a} and the tetra-quark \cite{fourq10,fourq14}. On the latter see  also the recent reviews \cite{EPP16},\cite{CCLZ16},\cite{ALS16},\cite{ARS17}.

Since   pentaquark and tetraquark systems, have already been found to exist one wonders  why not multi quark configurations in he nucleus? So we will examine the possibility of the presence of six quark clusters in the nucleus, a much more complex problem. Such clusters, if present with a reasonable probability in the nucleus, may contribute  to various proceses, like neutrinoless double beta decay mediated by heavy neutrinos or other exotic particles. In conventional nuclear physics the relevant nuclear matrix elements are suppressed due to the presence of the nuclear hard core.
In this presence of such clusters, however,   the interacting quarks are in the same hadron. So one can have a contribution even in the
case of of a $\delta$-function interaction \cite{JDV85}. Symmetries, of course, play a crucial role in reliably  estimating the probability of finding such six quark clusters in the nucleus.

In this article we develop the formalism  needed in the evaluation of the energy of all six-quark cluster configurations, which  can arise in a harmonic oscillator basis up to $2 \hbar \omega$ excitations. The symmetries that were found useful for this purpose were the combined spin color symmetry $SU_{cs}(6)$, the orbital symmetry $SU_{o}(6)$ and the isospin symmetry $SU_I(2)$.

\section{Basis States}

All states are products of four components: space, color, spin, and
isospin; the color and spin will be combined in color-spin wave
function described by $SU(6)_{cs}$.  We assume the total color of the six quarks
must be a color singlet.  (Note that this is a restriction: the only requirement is that the entire nucleus be a color singlet which leaves open the possibility that the six quarks and the remaining $A-2$ nuclear system could be non-singlet states while coupling to a color singlet state.  However, shell model codes that would be used to described the $A-2$ system currently assume the nucleons are in color singlet states so we make this assumption here.)   The isospin will be described by SU(2) and
the spatial wave functions -- being three-dimensional harmonic
oscillator functions -- will be labelled by $S(N), [f]_r$, and
$SU(3)_r, (\lambda \mu)_r$.  The total wave function is necessarily
antisymmetric.  This imposes restrictions on the possible $SU(6)_{cs}$
and $SU(2)_I$ representations.

The $SU(3)$ representations, be they flavour, color, or spatial, are
labelled by $(\lambda \mu)$.  The dimension of a $(\lambda \mu)$ represntation is
$\frac{1}{2}(\lambda + 1)(\mu + 1)(\lambda + \mu + 2)$.  The labels of
the $SU(2)$ and $SU(1)$ subgroups are:

\begin{center}
%\begin{table}\centering
\begin{tabular}{cclr}
$Y$&=&$p+q-\frac{1}{3}(2\lambda + \mu)$&\hspace{.5in}$0 \le
p\le\lambda$\\\\
$I$ &=&$ \frac{1}{2}(\mu + p - q)$&\hspace{.5in}$0 \le q\le\mu$\\\\
$I_3 $&=& $r-I$&$0 \le r\le\mu + p - q$\\\\
\end{tabular}
%\end{table}
\end{center}
Alternatively, the number of each of the three types of quarks for
color/flavor/spatial $SU(3)$ is:

%\begin{center}

\begin{eqnarray}
n_1 /n_u /n_z&=&\frac{1}{3}(N-\lambda)+r-\frac{2}{3}\mu +
q\nonumber\\\nonumber \\
n_2 / n_d / n_x &=& \frac{1}{3}(N-\lambda)-r+\frac{1}{3}\mu + p\\
n_3 / n_s/n_y &=& \frac{1}{3}(N+2\lambda+\mu)-p -
q\nonumber\\\nonumber\\\nonumber
n_u + n_d&=&\frac{1}{3}(2N-2\lambda-\mu)+p +
q\nonumber\\\nonumber\\\nonumber
\label{eq:defns}
\end{eqnarray}
where $N$ is the total number of quarks.
Thus, for the $(0\;0)$ representation, one has an equal number of
each type of quark, as one should.

Below we will construct wavefunctions using $SU(2)_I$ rather than
$SU(3)_f$.  We now demonstrate that
we can do this with no loss of generality {\it if} we assume there
are no strange quarks in the wave 
function.  From Eq.~\ref{eq:defns} above we see that the requirement
for zero $s$ quarks is that
$$n_u + n_d=\frac{1}{3}(2N-2\lambda-\mu)+p + q = N.$$
Taking $p$ and $q$ to be their maximum values, we have
$$N = n_u + n_d = \frac{1}{3}(2N+\lambda+2\mu) $$
or $N=\lambda+2\mu$.  Since the total number of quarks in a
$SU(3)_{f}$ representation $[f_1f_2f_3]$
 (which is 
contragradeint to $SU(6)_{cs}$) is
$$N_{tot} = f_1 + f_2+f_1 = \lambda + 2\mu + 3f_3,$$
we have the requirement that $f_3 = 0$.  Thus, we can label the
representations by either $SU(3)_f$ or 
$SU(2)_I$.

\subsection{$0\hbar \omega$ Excitations}

\begin{table}\centering
\begin{tabular}{|c|c|c|}
\hline
$[f]_{cs} $ & S & I\\\hline
$[42]_{cs}$ & 1&NA\\
$[411]_{cs}$&0&NA\\
$[33]_{cs}$&0&NA\\
$[321]_{cs}$&1, 2&NA\\
$[3111]_{cs}$&1&NA\\\hline
$[222]_{cs}$&1, 3&0\\
$[2211]_{cs}$&0, 2&1\\
$[21^4]_{cs}$&1&2\\
$[1^6]_{cs}$&0&3\\
\hline
\end{tabular}\caption[Six-quark $SU(6)_{cs}$ representations that
have color singlets]{The six-quark $SU(6)_{cs}$ representations that
have color singlet states, $(0 0)_c$, the 
allowed spin  and their concomitant isospin; these can be found from
Ref. \cite{dds1}.  $SU(6)_{cs}$ representations with more than two
columns
are not allowed.}
\label{tab:6qIR}
\end{table}

The $SU(6)_{cs}$ representations that contain a color singlet and
their spin content are shown in Table 1.  All six quarks are assumed
to be in the $0s$ state.  The wave function is then
\begin{equation}
| (0s)^6 [6]_r\; [f]_{cs}(0 0)_c \;S \;I=1>\; .\label{eq:basis0s6}
\end{equation}
Since the spatial wave function is totally symmetric, the product of
the color-spin and isospin wave functions must be antisymmetric.  For
$I=1$, the isospin representation is necessarily$[42]_I$.  The
conjugate representation is $[2211]_{cs}$; from Table~\ref{tab:6qIR}
one sees the spin can be either 0 or 2.  The former is diproton-like
but in the nucleus, one cannot exclude the S=2 possibility.

Note that the six quarks can also have isospin two or three as well as
one or two.  One may think of these states as having $\Delta$
admixtures.  However, even if one decomposes the $I=0$ six-quark wave
function into two hadrons, there will also be $\Delta$ admixtures.

\subsection{One Particle, $2\hbar\omega$ Excitations}

The wave function for positive parity is 
\begin{equation}
| (0s)^5 [5]_r\; [f]_{cs}(0 1)_c\; L_1=0\;\;S_1 \;I_1 \;\times\;
(1s0d)_r [1]_{cs}(1 0)_c\; S_2=\frac{1}{2}\; \;I_2=\frac{1}{2};\; (0
0)_c\; L=L_2\;S \;J\;I=1>.\label{eq:basis0s5}
\end{equation}
Since the $0s$ quarks and the $(1s0d)$ quarks are distinguishable, one
need not antisymmetrize the entire wave function.  The five-quark wave
function then can have either $I_1=\frac{1}{2}$ or $I_1=\frac{3}{2}$
if the total isospin of the six quarks is 1.  The five-quark states
that have a $SU(3)_c$ representation of $(0 1)_c$ and their spin and
isospin content are listed in Table~\ref{tab:5qIR}.

\begin{table}\centering

\begin{tabular}{|c|c|c|}
\hline $[f_{cs}]$ & $(\lambda \mu )_c \;S$& I\\\hline
\hline $[41]_{cs}$ & $(0 1)_c \;\frac{1}{2}$& NA\\
\hline $[32]_{cs}$ & $(0 1)_c \;\frac{1}{2}, \frac{3}{2}$&NA\\
\hline $[311]_{cs}$ & $(0 1)_c \;\frac{1}{2}, \frac{3}{2}$&NA\\
\hline &&\\$[221]_{cs}$ & $(0 1)_c \;\frac{1}{2}, \frac{3}{2},
\frac{5}{2}$&$\frac{1}{2}$\\&&\\
\hline &&\\$[2111]_{cs}$ & $(0 1)_c \;\frac{1}{2},
\frac{3}{2}$&$\frac{3}{2}$\\&&\\
\hline &&\\$[1^5]_{cs}$ & $(0 1)_c \;\frac{1}{2}$&$\frac{5}{2}$\\&&\\
\hline\end{tabular}
\caption[Five-quark $SU(6)_{cs}$ representations  that contain $(0
1)_c$]{ $SU(6)_{cs}$ representations for five quarks that contain $(0
1)_c$, their spin, and isospin; 
these can be found from Ref. \cite{dds1}.  Note that those
$SU(6)_{cs}$ representations that have more than two columns are not
allowed because the contragradient representation must belong to
$SU(2)_I$.
}
\label{tab:5qIR}
\end{table}%\nopagebreak

If the isospin of the five quarks, $I_1$, is $\frac{1}{2}$, the SU(2)
representation is $[32]_I$; the conjugate $SU(6)_{cs}$ is then
$[221]_{cs}$ which allows $S_1 = \frac{1}{2},\frac{3}{2},\frac{5}{2}.$ If the
isospin, $I_1$, is $\frac{3}{2}$, the SU(2) representation is
$[41]_I$; the conjugate $SU(6)_{cs}$ is then $[2111]_{cs}$ which
allows $S_1 = \frac{1}{2},\frac{3}{2}.$ The angular momentum, $L_2$,
of the $(1s0d)$ quark can be either 0 or 2.
The possible states are listed in Table~\ref{tab:s5p}.  These states 
can combine to form a variety of total $L, S, J, I$.

\begin{table}\centering

\begin{tabular}{|c|c|}
\hline &\\
 $(0s)^5 [5]_r\; [221]_{cs}(0 1)_c \;L_1=0\;\;S_1=\frac{1}{2} \;\;I_1
=\frac{1}{2}$& $(1s0d)_r [1]_{cs}(1 0)_c
\;  \;L_2\; S_2=\frac{1}{2}\;\;I_2=\frac{1}{2}$\\&\\
$ (0s)^5 [5]_r\; [221]_{cs}(0 1)_c \;L_1=0\;\;S_1=\frac{3}{2} \;\;I_1
=\frac{1}{2}$& $(1s0d)_r [1]_{cs}(1 0)_c
\;  \;L_2\; S_2=\frac{1}{2}\;\;I_2=\frac{1}{2}$\\&\\
$ (0s)^5 [5]_r\; [221]_{cs}(0 1)_c \;L_1=0\;\;S_1=\frac{5}{2} \;\;I_1
=\frac{1}{2}$& $(1s0d)_r [1]_{cs}(1 0)_c
\;  \;L_2\; S_2=\frac{1}{2}\;\;I_2=\frac{1}{2}$\\&\\
\hline&\\
$ (0s)^5 [5]_r\; [2111]_{cs}(0 1)_c \;L_1=0\;\;S_1=\frac{1}{2}
\;\;I_1 =\frac{3}{2}$&$ (1s0d)_r [1]_{cs}(1 0)_c
\;\;L_2\; S_2=\frac{1}{2}  \;\;I_2=\frac{1}{2}$\\&\\
$ (0s)^5 [5]_r\; [2111]_{cs}(0 1)_c \;L_1=0\;\;S_1=\frac{3}{2}
\;\;I_1 =\frac{3}{2}$&$ (1s0d)_r [1]_{cs}(1 0)_c
 \;\;L_2\; S_2=\frac{1}{2}\; \;I_2=\frac{1}{2}$\\&\\
\hline\end{tabular}
\caption[Basis states for $0s^5 (1s0d)$]{Basis states for $0s^5
(1s0d)$}
\label{tab:s5p}
\end{table}

\subsection{Two-Particle, $2\hbar\omega$ Excitations}

The wave function for positive parity states is
\begin{equation}
| (0s)^4 [4]_r\; [f]_{cs}(\lambda_c \mu_c)_c\; L_1=0\;\;S_1 \;I_1
\;\times\; (0p)_r^2\;[f_r] (\lambda_r \mu_r)\;
\;[f_2]_{cs}\;(\mu_c\lambda_c)_c\; S_2\; \;I_2;\; (0 0)_c\; L\;S
\;J\;I=1>.\label{eq:basis0s4}
\end{equation}
Again, since the $0s$ quarks and the $0p$ quarks are distinguishable,
one need not antisymmetrize the entire wave function.  The $0p^2$ wave
function can have a $SU(6)_{cs}$ representation of either $[2]_{cs}$
or $[11]_{cs}$.

\subsection{$0p^2$ States}
We have an additional feature for the $0p^2$ wave functions, namely
the spatial symmetry $[f_r]$ may be either [2] or [11].  We must have
\begin{equation}
|space \times SU(6)_{cs} \times isospin >
\end{equation}
totally antisymmetric.  For two quarks this is easily handled - if we
had three of more quarks, one would need resort to more sophisticated
results from the symmetric group.  Using obvious notation one has the
four possible states: $|S S A>, |S A S>, |A S S>,$ and $ |A A A>$.
The spatial wave functions are $[2]_r (2 0)$ and $[11]_r (0 1)$ -
since in our cases, the $SU(3)_{r}$ representations are determined by
$[f]_{r}$, either the $[f]_{r}$ or $SU(3)_{r}$ label is redundant.
The wave functions belonging to $[2]_{r}$ are spatially symmetric and
$[11]_{r}$ antisymmetric.  {\sl Note that with spatial symmetry
included and which can be either symmetric or anti-symmetric,
$[f]_{cs}$ no longer determines $]f]_{f}$ uniquely.}

The SU(6) color-spin representations are for $[2]_{cs}$ $(2 0)_c
\;S=1$ and $(0 1)_c \;S=0$ ({\it i.e.}, $|S S>$ or $|A A>$) and for
$[11]_{cs}$, $(2 0)_c \,S=0$ and $(0 1)_c\,S=1$.  If $(\lambda_{c}
\mu_{c} )_{c}$ is the $SU(3)_c$ representation, then $(-)^{\mu_{c} +
S}$ is -1 for the symmetric representations and 1 for the
antisymmetric representations.

One is familiar with the rule for two nucleons that $L + S + I$ must
be odd.  We can generalize the rule to $$ (-)^{L + S + \mu_{c} + I}$$
must be even.

The possible two-quark states are thus $\Big|[2]_{r}\,[2]_{cs}
(0\,1)_{f}\,I=0\,\big>$, $\Big|[2]_{r}\,[11]_{cs}
(2\,0)_{f}\,I=1\,\big>$, $\Big|[11]_{r}\,[2]_{cs}
(0\,1)_{f}\,I=1\big>$, and $\Big|[11]_{r}\,[11]_{cs}
(2\,0)_{f}\,I=0\big>$.

\subsection{$0s^4$ States}

Again, the spatial part is symmetric so we need deal only with
color-spin and isospin.  The color SU(3) representation will be the
contragradient of the color representation of the $(0p)^2$ particles
and will be either $(0 2)_c$ or $(1 0)_c$.  The possible SU(2) isospin
representations are $[22] I=0$, $[31] I=1$, and $[4] I=2$.  The
corresponding $SU(6)_{cs}$ representations are $[22]_{cs},
[211]_{cs}$, and $[1111]_{cs}$, respectively.  The color-spin content
is shown in Table~\ref{tab:4qIR}.

\begin{table}\centering

\begin{tabular}{|c|c|c|}
\hline $[f_{cs}]$ & $(\lambda \mu )_c \;S$& I\\
\hline $[22]_{cs}$ & (0 2) 0 & 0\\
 & (0 2) 2& 0\\
& (1 0) 1& 0\\\hline
$[211]_{cs}$ & (0 2) 1& 1\\
 & (1 0) 0& 1\\
 & (1 0) 1& 1\\
 & (1 0) 2& 1\\\hline
$[1^4]_{cs}$ & (0 2) 0& 2\\
 & (1 0) 1& 2\\
\hline \end{tabular}
\caption[Four-quark $SU(6)_{cs} $ represetations  that contain $(0
\;2)_c$ or $(1\; 0)_c$]{$SU(6)_{cs} $ represetations for four quarks
that contain $(0 \;2)_c$ or $(1\; 0)_c$, and 
the concomitant spin and isospin.}
\label{tab:4qIR}
\end{table}

\section{ Two-Quark Matrix Elements}

\subsection{Enumeration of Positive Parity Two-Quark States}
For the $0\hbar\omega$ one has the states listed in Table \ref{tab:2qpos}.

\begin{table}\centering
\begin{tabular}{|c|c|c|c|c|}
\hline
& $q^N$&$(\lambda \mu)_c$&I&L\\
\hline
1&$0s^2$&$$[2] (2 0) S=1$$&0&0\\
2& $0s^2$&[2] (0 1) S=0&0&0\\
3&$0s^2$& [11] (2 0) S=0&1&0\\
4& $0s^2$&[11] (0 1) S=1&1&0\\
\hline
5&0s\,1s&(2 0$)_c\;$S=0&0&0\\
6&0s\,0d&(2 0$)_c\;$S=0&0&2\\
7&0s\,1s&(0 1$)_c\;$S=1&0&0\\
8&0s\,0d&(0 1$)_c\;$S=1&0&2\\
9&0s\,1s&(2 0$)_c\;$S=0&1&0\\
10&0s\,0d&(2 0$)_c\;$S=0&1&2\\
11&0s\,1s&(0 1$)_c\;$S=1&1&0\\
12&0s\,0d&(0 1$)_c\;$S=1&1&2\\
\hline
13&$0p^2$&(0 1$)_c\;$S=0&0&0\\
14&$0p^2$&(2 0$)_c\;$S=1&0&0\\
15&$0p^2$&(0 1$)_c\;$S=1&1&0\\
16&$0p^2$&(2 0$)_c\;$S=0&1&0\\
17&$0p^2$&(0 1$)_c\;$S=0&0&1\\
18&$0p^2$&(2 0$)_c\;$S=1&0&1\\
19&$0p^2$&(0 1$)_c\;$S=1&1&1\\
20&$0p^2$&(2 0$)_c\;$S=0&1&1\\
21&$0p^2$&(0 1$)_c\;$S=0&0&2\\
22&$0p^2$&(2 0$)_c\;$S=1&0&2\\
23&$0p^2$&(0 1$)_c\;$S=1&1&2\\
24&$0p^2$&(2 0$)_c\;$S=0&1&2\\
\hline
\end{tabular}
\caption[Two-Quark States, Even Parity]{Two-Quark States, Even Parity}
\label{tab:2qpos}
\end{table}

\begin{table}\centering
\begin{tabular}{|c|c|c|c|c|c|}
\hline
& $q^N$&$(\lambda \mu)_c$&I&L&J\\
\hline
1&$0s\,0p$&$$ (2 0) S=1$$&0&1\\
2& $0s\,0p$& (0 1) S=0&0&1\\
3&$0s\,0p$&  (2 0) S=0&1&1\\
4& $0s\,0p$& (0 1) S=1&1&1\\
\hline
\end{tabular}
\caption[Two-Quark States, Odd Parity]{Two-Quark States, Odd Parity}
\label{tab:2qneg}
\end{table}

\subsection{Enumeration of Negative Parity Two-Quark States}
For the $0\hbar\omega$ one has the states listed in Table \ref{tab:2qneg}.

\section{ Matrix Elements of the Two-Quark Interaction}

Since the quark-quark interaction is two-body, we must separate off
two quarks from the remaining four.  We will use the same techniques 
as in the nuclear shell model\cite{Talmi} and in the MIT bag 
paper\cite{dds2}.

%\end{document}

%\subsection{Zero $\hbar \omega$ Excitations}
\subsection{$<0s^6\;  |V|\, 0s^6\;  >$}
This is the simplest case.  Two-body matrix elements for $0s^5$ and
$0s^4$ will be similar except for the labels.  Recall that $S$ in the
basis state of Eq.~(\ref{eq:basis0s6}) can be either 0 or 2.  
We take first the $J=S=0$ case.

For the six-quark color-spin symmetry $[2211]_{cs}$ one may have 
$[22]_{cs}\times[11]_{cs},\; [211]_{cs}\times[11]_{cs}, 
\;[1111]_{cs}\times[11]_{cs},$ and $[211]_{cs}\times[2]_{cs}$.

\begin{eqnarray}
&&| (0s)^6\, [6]_r\,(0 0)_r\; [2211]_{cs}(0 0)_c \;S=0 \;I=1\;J=0>
\;\;=\; \nonumber\\
&& \sum_{f'',(\lambda'',\mu''),I'',[f_2],I_2}\sqrt{\frac{n_{f''}}{n_{[2211]}}}
\;\;\Bigg\langle\begin{array}{cc}[f'']_{cs}  & [f_2]_{cs}\\
(\lambda'' \mu'')_c \,S'' &(\lambda_2 \mu_2)_c\,S_2\\ 
\end{array}\Bigg|\begin{array}{c}[2211]_{cs}\\(0 0)_c \;S \end{array}
\Bigg\rangle 
\Bigg\langle\begin{array}{cc}[4]_r  & [2]_r \\(0 0)_r &(0 0)_r
\nonumber\\ \end{array}  \Bigg |
\begin{array}{c}[6]_r\\(0 0)_r \end{array}  \Bigg\rangle \\ &&
\hspace{2.5cm}\times\;\;\;\Bigg\langle\begin{array}{cc}[\tilde{f''}]_I
& [\tilde{f}_2]_I\\I'' &I_2\\ \end{array}\Bigg|
\begin{array}{c}[42]_I\\I \end{array} \Bigg\rangle
\\
&\times& \Bigg[\Big| (0s)^4\, [4]_r\,(0 0)_r\;
[f'']_{cs}(\lambda''\mu'')_c \;S'' \;I''\,\Big> 
\times \;\Big| (0s)^2\, [2]_r\,(0 0)_r\; [f_2]_{cs}(\lambda_2
\mu_2)_c \;S_2 \;I_2 \Big>
\Bigg]^{(L,S,I,J)}\;.\label{eq:first}\nonumber
\end{eqnarray}

The first factor is the ratio of the dimensions of the representation
$[f'']$ and $[2211]$ of the symmetric group.  The next three factors
on the right-hand side of Eq~({\ref{eq:first}) are the Clebsch-Gordan
coefficients for the groups $SU(6)_{cs}, SU(3)_r,$ and $SU(2)_I$,
respectively.  We also use the notation that $\big[\;\;\big]^{L, S,
I, J}$ denotes the angular-momentum coupling of the relevant quantites
to $L, S, I, J$.  In this simple case, all orbital angular momentum is
zero, so $L=0$ and $S=J$.  Also, the $\tilde{f}$ indicates the
representation $[\tilde{f}]$ is contragradient to that of $[f]$.

The first of these factors can be obtained using the results of Ref.
\cite{dds2}; the relevant coefficients are shown in
Table~\ref{tab:6q2211}.  Since all quarks are in a $0s$ state, the
second factor, $\Bigg\langle\begin{array}{cc}[4]_r & [2]_r \\(0 0)_r
&(0 0)_r \\
\end{array} \Bigg | \begin{array}{c}[6]_r\\(0 0)_r \end{array}
\Bigg\rangle$, is necessarily of unit magnitude.  We can without loss
of generality assume it to be one.  Finally, the coefficient
$\Bigg\langle\begin{array}{cc}[\tilde{f''}]_I & [\tilde{f}_2]_I\\I''
&I_2\\ \end{array}\Bigg| \begin{array}{c}[42]_I\\I \end{array}
\Bigg\rangle$ describes how the $SU(2)_I$ representation $[42]_I$ with
angular momentum $I$ can be decoupled into representations
$[\tilde{f}'']_I$ and $[\tilde{f}_2]_I$.  Since there is a one-to-one
correspondence in $SU(2)$ between $[f]$ and the angular momentum $I$,
thse coeficients are all one; any other isospin dependence is in the
Clebsch-Gordan coefficients summarized in $[...]^I$.

\begin{table}[h]
\centering

\begin{tabular}{|c|c|}\hline
    [2]  &1 \\\hline
    [11]  &1 \\\hline\hline
    [21]  &2 \\\hline\hline
    [22]  &2 \\\hline
    [211] &3 \\\hline
    [1111]&1 \\\hline\hline
    [221]  &5 \\\hline
    [2111] &4 \\\hline
    [11111]&1 \\\hline\hline
    [222]  &5 \\\hline
    [2211]  &9 \\
\hline\end{tabular}
\caption{ Dimensions of representations of $S_{n}\,$.}
\label{tab:dimens}

\end{table}

\begin{table}[h]
\centering

\begin{tabular}{|c|c|c|}\hline
    [22]  &$[21] $&$1 $  \\\vspace{-.3cm}&&\\\hline
    [211] &$[21] $&$\frac{1}{2} $  \\\vspace{-.3cm}&&\\\hline
    [211] &$[111] $&$\frac{1}{2} $  
    \\\vspace{-.3cm}&&\\\hline\hline
    [1111] &$[111] $&$1 $  
    \\\vspace{-.3cm}&&\\\hline\hline\hline
    [221]  &$[22] $&$\frac{2}{5} $  \\\vspace{-.3cm}&&\\\hline
    [221]  &$[211] $&$\frac{3}{5} $  \\\vspace{-.3cm}&&\\\hline\hline
    [2111] &$[211] $&$\frac{3}{4} $  \\\vspace{-.3cm}&&\\\hline
    [2111] &$[1111] $&$\frac{1}{4} $  
    \\\vspace{-.3cm}&&\\\hline\hline
    [11111] &$[1111] $&$1 $  
    \\\vspace{-.3cm}&&\\\hline\hline\hline
    [222]  &$[221] $&$1 $ \\\vspace{-.3cm}&&\\\hline\hline
    [2211]  &$[221] $&$\frac{5}{9} $ \\\vspace{-.3cm}&&\\\hline
    [2211]  &$[2111] $&$\frac{4}{9} $ 
    \\
\hline\end{tabular}
\caption{ Dimensions of representations of $S_{n}\,$.}
\label{tab:weights1}

\end{table}

\begin{table}[h]
\centering

\begin{tabular}{|c|c|c|}\hline
    [22]  &$[2] \times [2]$&$\frac{1}{2} $  \\\vspace{-.3cm}&&\\\hline
    [22]  &$[11] \times [11]$&$\frac{1}{2} $  
    \\\vspace{-.3cm}&&\\\hline\hline
    [211] &$[2] \times [11]$&$\frac{1}{3} $  \\\vspace{-.3cm}&&\\\hline
    [211] &$[11] \times [2]$&$\frac{1}{3} $  \\\vspace{-.3cm}&&\\\hline
    [211] &$[11] \times [11]$&$\frac{1}{3} $  
    \\\vspace{-.3cm}&&\\\hline\hline
    [1111] &$[11] \times [11]$&$1 $  
    \\\vspace{-.3cm}&&\\\hline\hline\hline
    [221]  &$[21] \times [2]$&$\frac{2}{5} $  \\\vspace{-.3cm}&&\\\hline
    [221]  &$[21] \times [11]$&$\frac{2}{5} $  \\\vspace{-.3cm}&&\\\hline
    [221]  &$[111] \times [11]$&$\frac{1}{5} $  
    \\\vspace{-.3cm}&&\\\hline\hline
    [2111] &$[21] \times [11]$&$\frac{1}{2} $  \\\vspace{-.3cm}&&\\\hline
    [2111] &$[111] \times [2]$&$\frac{1}{4} $  
    \\\vspace{-.3cm}&&\\\hline
    [2111] &$[111] \times [11]$&$\frac{1}{4} $  
    \\\vspace{-.3cm}&&\\\hline\hline
    [11111] &$[111] \times [11]$&$1 $  
    \\\vspace{-.3cm}&&\\\hline\hline\hline
    [222]  &$[22] \times [2]$&$\frac{2}{5} $ \\\vspace{-.3cm}&&\\\hline
    [222]  &$[211] \times [11]$&$\frac{3}{5} $ 
    \\\vspace{-.3cm}&&\\\hline\hline
    [2211]  &$[22] \times [11]$&$\frac{2}{9} $ \\\vspace{-.3cm}&&\\\hline
    [2211]  &$[211] \times [2]$&$\frac{3}{9} $ \\\vspace{-.3cm}&&\\\hline
    [2211]  &$[211] \times [11]$&$\frac{3}{9} $ \\\vspace{-.3cm}&&\\\hline
    [2211]  &$[1111]\times [11] $&$\frac{1}{9} $ 
    \\
\hline\end{tabular}

\caption{ Weight factors for two particles; ratios of the dimensions
of representations of $S_{n}\,$.}
\label{tab:weights2}
\end{table}

With these simplications we have

\begin{eqnarray}
&&| (0s)^6\, [6]_r\,(0 0)_r\; [2211]_{cs}(0 0)_c \;S=0 \;I=1\;J=0>\;
\\
= 
&&\sum_{f'',(\lambda'',\mu''),I'',[f_2],I_2}\sqrt{\frac{n_{f''}}{n_{[2211]}}}
\;\;\Bigg\langle\begin{array}{cc}[f'']_{cs}\;\;\;\;\;\;\;\;\;[f_2]_{cs}\\
(\lambda'' \mu'')_c \,S'' \;\;\; (\lambda_2 \mu_2)_c\,S_2\\
\end{array}\Bigg|\begin{array}{c}[2211]_{cs}\\(0 0)_c \;S \end{array}
\Bigg\rangle \nonumber\\
&\times& \Bigg[\Big| (0s)^4\, [4]_r\,(0 0)_r\;
[f'']_{cs}(\lambda''\mu'')_c \;S'' \;I''\,\Big> \times \;\Big|
(0s)^2\, [2]_r\,(0 0)_r\; [f_2]_{cs}(\lambda_2 \mu_2)_c \;S_2 \;I_2
\Big> \Bigg]^{(L,S,I,J)}\;.\label{eq:sec}
\end{eqnarray}

The allowed $[f'']_{cs}$ representations are from
Table~\ref{tab:6q2211}, $[22], [211],$ and $[1111]$.

The matrix elements of a two-body operator are then:

\begin{eqnarray}
&&\hspace{-.9cm}< (0s)^6\, [6]_r\,(0 0)_r\; [2211]_{cs}(0 0)_c \;S=0
\;I=1\;J=0\Big|\;\sum_{i,j}V_{ij}\;\Big|
(0s)^6\, [6]_r\,(0 0)_r\; [2211]_{cs}(0 0)_c \;S=0 \;I=1\;J=0>\;
\nonumber\\
&&\hspace{2.cm}= \frac{6\cdot
5}{2}\sum_{f'',(\lambda'',\mu''),I'',[f_2],I_2}
\frac{n_{f''}}{n_{[2211]}}
\;\;\Bigg\langle\begin{array}{cc}[f'']_{cs} & [f_2]_{cs}\\
(\lambda'' \mu'')_c \,S'' &(\lambda_2 \mu_2)_c\,S_2\\
\end{array}\Bigg|\begin{array}{c}[2211]_{cs}\\(0 0)_c \;S \end{array}
\Bigg\rangle \nonumber\\
 \;\;&&\hspace{2.cm}\times\;\;\Bigg\langle\begin{array}{cc}[f'']_{cs}&
[f_2']_{cs}\\
(\lambda'' \mu'')_c \,S'' &(\lambda_2' \mu_2')_c\,S_2'\\
\end{array}\Bigg|\begin{array}{c}[2211]_{cs}\\(0 0)_c \;S \end{array}
\Bigg\rangle \;\;\;\;
U\Bigg(\begin{array}{ccc}0&0&0\\S''&S_2&S\\J''&J_2&J\end{array}\Bigg)
\nonumber\\
&&\;\hspace{1.cm}\times\;\;\; \Big< (0s)^2\, [2]_r\,(0 0)_r\;
[f_2']_{cs}(\lambda_2' \mu_2')_c \;S_2' \;I_2 \Big| \;V_{56} \;\Big|
(0s)^2\, [2]_r\,(0 0)_r\; [f_2]_{cs}(\lambda_2 \mu_2)_c \;S_2 \;I_2
\Big>\;.\label{eq:first1}
\end{eqnarray}
Since all the orbital angular momentum is zero, the $9J$ symbol is one
and $J''=S''$, $J_2=S_2$, and $S=J$.  If the two-quark interaction
does not change spin, $S_2' = S_2$, then it follows that $(\lambda_2
\mu_2)_c = (\lambda_2' \mu_2')_c$.  Hence,

\begin{eqnarray}
&&\hspace{-.9cm}< (0s)^6 \; [2211]_{cs}\;(0 0)_c \;S=0
\;I=1\;J=0\;\Big|\;\sum_{i,j}V_{ij}\;\Big|\;
(0s)^6\, \; [2211]_{cs}\;(0 0)_c \;S=0 \;I=1\;J=0>\;
\nonumber\\
&&\hspace{2.cm}= \frac{6\cdot
5}{2}\sum_{f'',(\lambda'',\mu''),I'',[f_2],I_2}
\frac{n_{f''}}{9}
\;\;\Bigg\langle\begin{array}{cc}[f'']_{cs} & [f_2]_{cs}\\
(\lambda'' \mu'')_c \,S'' &(\lambda_2 \mu_2)_c\,S_2\\
\end{array}\Bigg|\begin{array}{c}[2211]_{cs}\\(0 0)_c \;S \end{array}
\Bigg\rangle \nonumber\\
 \;\;&&\hspace{2.cm}\times\;\;\Bigg\langle\begin{array}{cc}[f'']_{cs}&
[f_2']_{cs}\\
(\lambda'' \mu'')_c \,S'' &(\lambda_2' \mu_2')_c\,S_2'\\
\end{array}\Bigg|\begin{array}{c}[2211]_{cs}\\(0 0)_c \;S \end{array}
\Bigg\rangle \;\;\;\;
\nonumber\\
&&\;\hspace{1.cm}\times\;\;\; \Big< (0s)^2\, \;
[f_2]_{cs}(\lambda_2 \mu_2 )_c \;S_2 \;I_2 \;J_2\Big| \;V_{56} \;\Big|
(0s)^2\, \; [f_2]_{cs}(\lambda_2 \mu_2)_c \;S_2 \;I_2\;J_2
\Big>\;.\label{eq:sixq-2}
\end{eqnarray}
In eq.  \ref{eq:sixq-2} it is explicitly assumed that the two-quark
interaction does not change $[f_{2}], (\lambda_2 \mu_2)_c, \;S_2,$ or
$I_2$.  Using the results from Table \ref{tab:6q2211} we have

\begin{eqnarray}
&&\hspace{-.9cm}< (0s)^6 \; [2211]_{cs}\;(0 0)_c \;S=0
\;I=1\;J=0\;\Big|\;\frac{1}{15}\sum_{i,j}V_{ij}\;\Big|\;
(0s)^6\, \; [2211]_{cs}\;(0 0)_c \;S=0 \;I=1\;J=0>\;
\nonumber\\
    &=&\frac{2}{9}\bigg[\frac{3}{4}<(2\; 0)_{c}\;S_{2}=0>+
        \frac{1}{4}<(0\; 1)_{c}\;S_{2}=1>\bigg]\nonumber\\
   &+&\frac{3}{9}\bigg[\frac{1}{2}<(2\; 0)_{c}\;S_{2}=1>+
         \frac{1}{2}<(0\; 1)_{c}\;S_{2}=0>\bigg]\nonumber\\
    &+&\frac{3}{9}\bigg[
                      <(0 \;1)_{c}\;S_{2}=1>\bigg]
   \; +\;\frac{1}{9}\bigg[\frac{3}{5}<(2\; 0)_{c}\;S_{2}=0>+
       \frac{2}{5}<(0 \;1)_{c}\;S_{2}=1>\bigg]\nonumber\\
     &=&\frac{7}{30}<(2\; 0)_{c}\;S_{2}=0>+
        \frac{13}{30}<(0\;1)_{c}\;S_{2}=1>+\frac{1}{6}<(2\;
	0)_{c}\;S_{2}=1>+
	\frac{1}{6}<(0\; 1)_{c}\;S_{2}=0>\;\;\;\;\;\;\;\;
\end{eqnarray}
where 
\begin{eqnarray}
  <(\lambda_2 \mu_2)_{c}\;S_{2}>=  \Big< (0s)^2\, \;
[f_2]_{cs}(\lambda_2 \mu_2 )_c \;S_2 \;I_2 \;J_2\Big| \;V_{56} \;\Big|
(0s)^2\, \; [f_2]_{cs}(\lambda_2 \mu_2)_c \;S_2 \;I_2\;J_2
\Big>.\nonumber
\end{eqnarray}
    
\subsection{$<0s^n\;  |V|\, 0s^n\; >$}
 
These matrix elements will be used for $n=4,5,6$ so we give the 
general formula here:
\begin{eqnarray}
&&| (0s)^n\, [n]_r\,(0 0)_r\; [f]_{cs}(\lambda_{c}\mu_{c})_c \;L_1\;S_1 \;I_1\;J_1>
\;\;=\; \nonumber\\
&& \sum_{f'',(\lambda_{c}'',\mu_{c}''),I'',[f_2],I_2}\sqrt{\frac{n_{f''_{cs}}}{n_{f_{cs}}}}
\;\;\Bigg\langle\begin{array}{cc}[f'']_{cs}  & [f_2]_{cs}\\
(\lambda_{c}'' \mu_{c}'')_c \,S'' &(\lambda_2 \mu_2)_c\,S_2\\ 
\end{array}\Bigg|\begin{array}{c}[f]_{cs}\\(0 0)_c \;S_1 \end{array}
\Bigg\rangle 
\;
\nonumber\\
&\times& \Bigg[\Big| (0s)^{n-2}\, [f''_{r}]_r\,(0 0)_r\;
[f_{cs}'']_{cs}(\lambda_{c}''\mu_{c}'')_c \;S'' \;I''\,\Big> 
\times \;\Big| (0s)^2\, [2]_r\,(0 0)_r\; [f_2]_{cs}(\lambda_2
\mu_2)_c \;S_2 \;I_2 \Big>
\Bigg]^{(L_1,S_1,I_1,J_1)}\;\nonumber   
\end{eqnarray}

Using 2B CFPs for an n-quark system:
\begin{eqnarray}
&&< q^N [f]_{cs} (\lambda \mu)_{cs}\;L S J \;  \Big| \sum_{i\le j} V_{i,j} \;\Big| \; q^N  [f']_{cs} (\lambda' \mu')_{cs}\,L'\,S'\,J\,> = \frac{1}{2} N(N-1)\sum_{f''_{cs}}\frac{n_{f''_{cs}}}{\sqrt{n_{f_{cs}}n_{\bar{f}_{cs}}}}\times\nonumber\\
&& 
\times\sum _{f_{12},(\lambda'', \mu'')\cdots}  \Bigg\langle\begin{array}{cc}[f'']_{cs}  & [f_{12}]_{cs}\\
(\lambda'' \mu'')_{cs} \,S'' &(\lambda_{12} \mu_{12})_{cs}\,S_{12}\\ 
\end{array}\Bigg|\begin{array}{c}[f]_{cs}\\(\lambda \mu)_{cs} \;S 
\end{array}
\Bigg\rangle
\Bigg\langle\begin{array}{cc}[f'']_{cs}  & [f_{12}']_{cs}\\
(\lambda'' \mu'')_{cs} \,S'' &(\lambda_{12}' \mu_{12}')_{cs}\,S'_{12}\\ 
\end{array}\Bigg|\begin{array}{c}[f']_{cs}\\(\lambda' \mu')_{cs} \;S '
\end{array}
\Bigg\rangle\nonumber\\
&&\times\;
\nonumber U \left(
\begin{array}{ccc}
L''&L_{12}&L\\
S''&S_{12}&S\\J''&J_{12}&J
\end{array}\right)\;U \left(
\begin{array}{ccc}
L''&L'_{12}&L'\\
S''&S_{12}'&S'\\J''&J_{12'}&J
\end{array}\right)\nonumber\\&& \times
\Big< \;q^{N-2} [f'']_{cs}\, (\lambda'' \mu'')_{cs} \,L''S'' J''\times q^2\,[f_{12}]_{cs}\;(\lambda_{12}\, \mu_{12})_{cs}\,L_{12}S_{12}\,, \,J_{12}\,\Big|\; V_{12}\,\Big| \nonumber\\
&&\hspace{3cm}\Big| \; q^{N-2}[f'']_{cs}\, (\lambda'' \mu'')_{cs} \,L''S'' J''\times q^2[f'_{12}]_{cs}\;(\lambda'_{12}\, \mu'_{12})_{cs}\,L'_{12}S'_{12}\, J_{12}\,
\Big>\nonumber
\end{eqnarray}

The $q^{N-2}$ part integrates to one and the remaining two-body matrix element is
\begin{eqnarray}
&&\Big<  q^2\,[f_{12}]_{cs}\;(\lambda_{12}\, \mu_{12})_{cs}\,L_{12}S_{12}\,\,J_{12}\,\Big|\; V_{12}\,\Big|  \;  q^2[f'_{12}]_{cs}\;(\lambda'_{12}\, \mu'_{12})_{cs}\,L'_{12}S'_{12}\,,\,J_{12}\,
\Big>\nonumber\\&&=(-)^{J_{12}-L_{12}+S_{12}'}W(L_{12}L_{12}'S_{12}S_{12}' ; k J_{12})\;\Big< L_{12}\,\Big|\Big|\,V_{12}^{(k)}\,\Big|\Big|\,L_{12}'\Big>\;\Big< S_{12}\,\Big|\Big|V_{12}^{(k)}\,\Big|\Big|\,S_{12}'\Big>
\nonumber\end{eqnarray}

But all the orbital angular momenta for the $0s$ quarks are zero so
$$
 U \left(
\begin{array}{ccc}
L''&L_{12}&L\\
S''&S_{12}&S\\J''&J_{12}&J
\end{array}\right) =  U \left(
\begin{array}{ccc}
0&0&0\\
S''&S_{12}&S\\J''&J_{12}&J
\end{array}\right) = 1\,,
$$
$$W(LL'SS' ; k J)= W(0 0 S S' ;0J) = \frac{1}{\sqrt{S}}\delta_{S S'}\delta_{S J}\;,
$$
and the matrix element for $q^N$ becomes

\begin{eqnarray}
&&< q^N [f]_{cs} (\lambda \mu)_{cs}\;L S J \;  \Big| \sum_{i\le j} V_{i,j} \;\Big| \; q^N  [f']_{cs} (\lambda' \mu')_{cs}\,L'\,S'\,J\,> = \frac{1}{2} N(N-1)\delta_{LL'}\delta_{SS'}\delta_{L0}\delta_{S J}\sum_{f''_{cs}}\frac{n_{f''_{cs}}}{\sqrt{n_{f_{cs}}n_{\bar{f}_{cs}}}}\times\nonumber\\
&& 
\times\sum _{f_{12},(\lambda'', \mu'')\cdots}  \Bigg\langle\begin{array}{cc}[f'']_{cs}  & [f_{12}]_{cs}\\
(\lambda'' \mu'')_{cs} \,S'' &(\lambda_{12} \mu_{12})_{cs}\,S_{12}\\ 
\end{array}\Bigg|\begin{array}{c}[f]_{cs}\\(\lambda \mu)_{cs} \;S 
\end{array}
\Bigg\rangle
\Bigg\langle\begin{array}{cc}[f'']_{cs}  & [f_{12}']_{cs}\\
(\lambda'' \mu'')_{cs} \,S'' &(\lambda_{12}' \mu_{12}')_{cs}\,S_{12}\\ 
\end{array}\Bigg|\begin{array}{c}[f']_{cs}\\(\lambda' \mu')_{cs} \;S 
\end{array}
\Bigg\rangle\nonumber\\
&&\times\frac{1}{\sqrt{S_{12}}}\;\Big< L_{12}=0\,\Big|\Big|\,V_{12}^{(0)}\,\Big|\Big|\,L_{12}'=0\Big>\;\Big< S_{12}\,\Big|\Big|\,V_{12}^{(0)}\,\Big|\Big|\,S_{12}\Big>\;\delta_{S_{12} S'_{12}}\delta_{S J}\delta_{k0}.
\label{eq:sn-sn}
\end{eqnarray}

\subsubsection{Examples}
\begin{eqnarray}
&&< q^5 [221]_{cs} (0 1)_{c}\;L=0 S J \;  \Big| \sum_{i\le j} V_{i,j} \;\Big| \; q^5  [2111]_{c} (0 1)_{cs}\,L'=0\,S'\,J\,> = 10\,\delta_{SS'}\delta_{S J}\sum_{f''_{cs}}\frac{n_{f''_{cs}}}{\sqrt{5\cdot4}}\times\nonumber\\
&& 
\times\sum _{f_{12},(\lambda'', \mu'')\cdots}  \Bigg\langle\begin{array}{cc}[f'']_{cs}  & [f_{12}]_{cs}\\
(\lambda'' \mu'')_{c} \,S'' &(\lambda_{12} \mu_{12})_{c}\,S_{12}\\ 
\end{array}\Bigg|\begin{array}{c}[221]_{cs}\\(0 1)_{c} \;S 
\end{array}
\Bigg\rangle
\Bigg\langle\begin{array}{cc}[f'']_{cs}  & [f_{12}']_{cs}\\
(\lambda'' \mu'')_{c} \,S'' &(\lambda_{12}' \mu_{12}')_{c}\,S_{12}\\ 
\end{array}\Bigg|\begin{array}{c}[2111]_{cs}\\(0 1)_{c} \;S' 
\end{array}
\Bigg\rangle\nonumber\\
&&\times\langle\, [f_{12}]\,(\lambda'' \mu'') \,S'' \bigg| V \bigg| [f_{12}']\,(\lambda'' \mu'') \,S''\,\rangle.
\nonumber
\end{eqnarray}

Note that $S=J=S'$.  Acceptable $[f'']$ are $[2 1]$ and $[111]$.  If $[f_{12}] = [f'_{12}]$, from Table \ref{tab:weights2} we have as possible products: $[21]\times[11]$ and $[111]\times[11]$.

\vspace{1.cm}
%\newpage 
\subsection{$<0s^5\; 1s0d\, |V|\, 0s^5\; 1s0d >$}
%\paragraph{$<0s^5\}
The $0s^5$ states for both the bra and ket are necessarily $[5]_r\;(0
1)_r\;(0 1)_{c}$;  $S_5$ can be $\frac{1}{2},\frac{3}{2}$  and $I_5$ can be $\frac{1}{2},\frac{3}{2}$.  

\begin{eqnarray}
&&< (0s)^5\,  [f]_{cs} \;S_5 \;I_5 J_5 \times 1s0d\;\ell \;;(0 0)_c
\;L\;S\;J \Big|  \;\sum_{i,j}V_{ij}\;\Big|
(0s)^5\, \, [\bar{f}]_{cs} \;\bar{S_5} \;\bar{I_5} \times 1s0d \;\bar{\ell} \;\;;(0 0)_c
\;\bar{L}\;\bar{S}\;J>\; \nonumber\\
&&=\;< (0s)^5\,  [f]_{cs} \;S_5 \;I_5 \Big| \;\sum_{i,j}V_{ij}\;\Big|
(0s)^5\, \, [\bar{f}]_{cs} \;\bar{S}_5 \;\bar{I}_5 >\delta_{\ell\;\bar{\ell}}\; \;+ \nonumber\\
&&+< (0s)^5\,  [f]_{cs} \;S_5 \;I_5 J_5 \times 1s0d\;\ell \;;(0 0)_c
\;L\;S\;J\Big|\;V_{s\;p}\;\Big|
(0s)^5\, \, [\bar{f}]_{cs} \;\bar{S}_5 \;\bar{I}_5 \times 1s0d \;\bar{\ell} \;\; ;(0 0)_c
\;\bar{L}=\bar{\ell}\;\bar{S}\;J>\nonumber
\end{eqnarray}

For $(0s)^{5}$ one can only have $L_5 = 0$ so the 9J is one.
\begin{eqnarray}
&&<(0s)^5\, \,(0 0)_r\; [f]_{cs}(0 1)_c \;L_5\;S_5
\;I_5\;J_5\Big|\;V\;\Big| (0s)^5\, \,(0 0)_r\; [\bar{f}]_{cs}(0 1)_c \;\bar{L_5}=0\;\bar{S_5}
\;\bar{I_5}\;J_5> \; \nonumber\\
&=&10\hspace{-.9cm}\sum_{f'',(\lambda'',\mu''),I'',[f_2],I_2}
\frac{n_{f''_{cs}}}{\sqrt{n_{f_{cs}}n_{\bar{f}_{cs}}}}
\;\;\Bigg\langle\begin{array}{cc}[f'']_{cs}  & [f_2]_{cs}\\
(\lambda'' \mu'')_c \,S'' &(\lambda_2 \mu_2)_c\,S_2\\ 
\end{array}\Bigg|\begin{array}{c}[f]_{cs}\\(0 1)_c \;S_5 \end{array}
\Bigg\rangle 
\Bigg\langle\begin{array}{cc}[f'']_{cs}  & \bar{[f_2]}_{cs}\\
(\lambda'' \mu'')_c \,S'' &(\bar{\lambda_2 }\bar{\mu_2})_c\,\bar{S_2}\\ 
\end{array}\Bigg|\begin{array}{c}[\bar{f}]_{cs}\\(0 1)_c \;\bar{S_5} 
\end{array}
\Bigg\rangle 
\;\nonumber\\&&\hspace{1.5cm}
\times\Big< (0s)^2\, [2]_r\,(0 0)_r\; [f_2]_{cs}(\lambda_2
\mu_2)_c \;S_2 \;I_2 \Big|\;V\;\Big| (0s)^2\; [2]_r\;(0 0)_r\; [\bar{f_2}]_{cs}(\bar{\lambda_2}
\bar{\mu_2})_c \;\bar{S_2} \;\bar{I_2} \Big> 
\end{eqnarray}

The second term is more interesting

\begin{eqnarray}
&&\Big|(0s)^5\, \, [f]_{cs} \;S_5 \;I_5 \times 
1s0d\;\;;(0 0)_c
\;L\;S\;J>\nonumber\\
&=&\sum_{f',(\lambda',\mu'),S',I',[f_2],I_2}\sqrt{\frac{n_{f'_{cs}}}{n_{f_{cs}}}}\;
    \Bigg\langle\begin{array}{cc}[f']_{cs}  & [1]_{cs}\\
(\lambda' \mu')_c \,S' &(1 0)_c\,S_2\\ 
\end{array}\Bigg|\begin{array}{c}[f]_{cs}\\(0 1)_c \;S_5 
\end{array}
\Bigg\rangle\nonumber\\
&\times&\Bigg| \Bigg[ [f'_{cs}](\lambda'\mu')_c S' \times 0s(1 0)\Bigg]^{[f_{cs}](0 1)_cS_5}1s0d; (00)_c \,L\,S\;JI\,\rangle\nonumber
\end{eqnarray}

The ket on the right can be written as (dropping the bar for now)
\begin{eqnarray}
&&\sum_{(\lambda_{12}\mu_{12})S_{12}I_{12}} U\Big( (\lambda'\mu') (1 
0) (0 0) (1 0); (0 1) 
(\lambda_{12}\mu_{12})\Big)U\Big(S'\frac{1}{2}S\frac{1}{2};S_5S_{12}\Big)
U\Big(I'\frac{1}{2}I\frac{1}{2};I_5I_{12}\Big)\nonumber\\
&\times&\Big|(0s)^4\, \, [f']_{cs}\,(\lambda'\mu')\;S' \;I' \times \Bigg[0s\,
[1]_{cs}\, (1 0)_{c} 
\;\frac{1}{2}\,\frac{1}{2}\;\times
1s0d\;(1 0)_{c}\,l\;\frac{1}{2}\,\frac{1}{2}\Bigg]^{{S_{12}I_{12}(\lambda_{12}\mu_{12})\,
}}\,;(0 0)_c
\;L\;S\;J>\nonumber\\
&=&\hspace{-1.2cm}\sum_{(\lambda_{12}\mu_{12})S_{12}I_{12}J_{12}J'} 
\hspace{-.9cm}
U\Big( (\lambda'\mu') (1  0) (0 0) (1 0); (0 1) 
(\lambda_{12}\mu_{12})\Big)U\Big(S'\frac{1}{2}S\frac{1}{2};S_5S_{12}\Big)
U\Big(I'\frac{1}{2}I\frac{1}{2};I_5I_{12}\Big)U
    \Bigg(\begin{array}{ccc}0 & L_{12} &L\\
S' & S_{12}& S\\ J' & J_{12} & J
\end{array}\Bigg)\nonumber
\\
&\times&\Big|\,(0s)^4\, \, [f']_{cs}\,(\lambda'\mu')\;S' \;I' J'\times \Bigg[0s\,
[1]_{cs}\, (1 0)_{c} 
\;\frac{1}{2}\,\frac{1}{2}\;\times
1s0d\;(1 0)_{c}\,l\;\frac{1}{2}\,\frac{1}{2}\Bigg]^{{S_{12}I_{12}(\lambda_{12}\mu_{12})L_{12}\,J_{12}	}}\,;(0 0)_c
\;J>\label{eq:s5sd}\\
&=&
\hspace{-1.2cm}\sum_{(\lambda_{12}\mu_{12})S_{12}I_{12}J_{12}J'} 
\hspace{-.9cm}
\delta_{(\lambda'\mu') (1 \,
0)}\delta_{(0\, 1)(\lambda_{12}\mu_{12})}\; U\Big(S'\frac{1}{2}S\frac{1}{2};S_5S_{12}\Big)
U\Big(I'\frac{1}{2}I\frac{1}{2};I_5I_{12}\Big)\,
    (-)^{S_{12}+J-S'-L_{12}}\delta_{L_{12}L}\delta_{S'J'}\,W\big(S_{12}\,S\,J_{12}\,J;S'\,l\big)\nonumber
\\
&\times&\Big|\,(0s)^4\, \, [f']_{cs}\,(\lambda'\mu')\;S' \;I' J'\times \Bigg[0s\,
[1]_{cs}\, (1 0)_{c} 
\;\frac{1}{2}\,\frac{1}{2}\;\times
1s0d\;(1 0)_{c}\,l\;\frac{1}{2}\,\frac{1}{2}\Bigg]^{{S_{12}I_{12}(\lambda_{12}\mu_{12})L_{12}\,J_{12}	}}\,;(0 0)_c
\;J>\nonumber
\end{eqnarray}
where the results from Vergados  (but we'll just use the code)
\cite{Vergados} 
\begin{eqnarray}
    U\Big( (\lambda''\mu'') (1  0) (0 0) (1 0); (0 1) 
(\lambda_{12}\mu_{12})\Big) &=&\delta_{(\lambda''\mu'') (1 \,
0)}\nonumber
\nonumber\end{eqnarray}
which defines $\Delta_{(\lambda_{12} \mu_{12})}$, 
and from ref. \cite{Talmi}
\begin{eqnarray}
U  \Bigg(\begin{array}{ccc}0 & L_{12} &L\\
S' & S_{12}& S\\ J' & J_{12} & J
\end{array}\Bigg)&=& 
(-)^{S_{12}+J-S'-L_{12}}\delta_{L_{12}L}\delta_{S'J'}\,W\big(S_{12}\,S\,J_{12}\,J;S'\,L_{12}\big)
\nonumber\end{eqnarray}
have been used.

The matrix element becomes

\begin{eqnarray}
&&\Big<(0s)^5\, \, [f_5]_{c} \;S_5 \;I_5 \times 
1s0d\;\ell\;\;;(0 1)_c
\;L\;S\;J>\Big|\;\sum V_{ij}\;\Big|(0s)^5\, \, [\bar{f_5}]_{c} \;\bar{S}_5 \;\bar{I}_5 \times 
1s0d\;\;;(0 1)_c \;\bar{\ell}
\;\bar{L}\;\bar{S}\;J>\nonumber\\
&=&5\sum_{f',(\lambda',\mu'),S',I',[f_2],I_2}\sqrt{\frac{n_{f'}^2}
{n_{f_{5}} \bar{n}_{f_{5}}}}\;
    \Bigg\langle\begin{array}{cc}[f']_{c}  & [1]_{c}\\
(\lambda' \mu')_c \,S' &(1 0)_c\,\frac{1}{2}\\ 
\end{array}\Bigg|\begin{array}{c}[f_5]_{c}\\(0 1)_c \;S_5 
\end{array}
\Bigg\rangle
\Bigg\langle\begin{array}{cc}[f']_{c}  & [1]_{c}\\
(\lambda' \mu')_c \,S' &(1 0)_c\,\frac{1}{2}\\ 
\end{array}\Bigg|\begin{array}{c}[\bar{f_5}]_{c}\\(0 1)_c \;\bar{S}_5 
\end{array}
\Bigg\rangle\nonumber\\
&&\times\hspace{1.cm}U\Big(S'\frac{1}{2}S\frac{1}{2};S_5S_{12}\Big)\,
U\Big(S'\frac{1}{2}\bar{S}\frac{1}{2};\bar{S}_5\bar{S}_{12}\Big)\,
U\Big(I'\frac{1}{2}I\frac{1}{2};I_5I_{12}\Big)\,
U\Big(I'\frac{1}{2}I\frac{1}{2};\bar{I}_5I_{12}\Big)\;\nonumber\\&&
\times\hspace{1.cm}U\Big( (\lambda'\mu') (1  0) (0 0) (1 0); (0 1) 
(\lambda_{12}\mu_{12})\Big)U\Big( (\lambda'\mu') (1  0) (0 0) (1 0); (0 1) 
(\bar{\lambda}_{12}\bar{\mu}_{12})\Big)
\nonumber\\&&\times \hspace{1.cm}U
    \Bigg(\begin{array}{ccc}0 & \ell &\ell\\
S' & S_{12}& S\\ J' & J_{12} & J
\end{array}\Bigg)U
    \Bigg(\begin{array}{ccc}0 & \ell &\ell\\
S' & S_{12}& \bar{S}\\ J' & J_{12} & J
\end{array}\Bigg)\nonumber\\
\hspace{-.cm}&&\times\hspace{.95cm}
\Big< \; 0s \;\times\;
1s0d\;\,l\,;\;(\lambda_{12}\mu_{12})_{c}
\;\ell\;S_{12}\;J_{12}I_{12}\,\Big| V \Big|\, 0s\,
  \times\; 1s0d\;\,\bar{l}\;\,;(\bar{\lambda}_{12}\bar{\mu}_{12})_c
\;\bar{\ell}\;\bar{S}_{12}\;J_{12}I_{12}>
\end{eqnarray}
Note that the SU(3) Racah coefficients lead to the requirement
$$(\lambda'\mu') = (\mu_{23} \lambda_{23})=(\mu'_{23} \lambda'_{23})\;.$$
However, there is no requirement that $S_{12} = S_{12}'$.

\subsubsection{Isospin}

Consider the case in which the isospin of the five-quark state differs and consider only the interaction amongst the $0s^5$ states:

\begin{eqnarray}&&<s^5 \alpha I_L \times 1s0d\,\ell \; J I \; | V | \; s^5 \beta I_R \times 1s0d, \ell'\; J\,I\;>\nonumber\\
&=& \delta_{\ell\ell'}\sum <0s^3\alpha''\times 0s^2 \alpha_{2L}|\}0s^5 \alpha>
<0s^3\alpha''\times 0s^2 \alpha_{2R}|\}0s^5 \beta>\,\nonumber\\
&\times&\bigg[<0s^2\alpha_{2L} \, I_2 |V|\,0s^2 \alpha_{2R}\, I_2 ><0s^3\alpha''\,I_3\,|0s^3 \alpha'' \,I_3\,>\bigg]^{I}_{M_T}\nonumber
\end{eqnarray}

It is implcitly assumed the bra and ket states are each coupled to isospin $I$:
\begin{eqnarray}
\bigg[|\,0s^2 \alpha_{2R}\, I_2 M_2>|0s^3 \alpha'' \,I_3\,>\bigg]^{I}_{M_T}=
\sum_{M''_{T} M_2} C \begin{array}{ccc}I_2 & I_3&I\\
 M_2 & M''_{T} & M_T
\end{array}
\bigg|\,0s^2 \alpha_{2R}\, I_2 M_2>|0s^3 \alpha'' \,I_3\,M_T''>\label{eq:iso}
\end{eqnarray}
But the two-body matrix elements are independent of ${M_T}$ and the three-body overlap is one so one can sum over the $m_T$ components:
$$\sum_{M''_T M_2}C \begin{array}{ccc}I_2 & I_3&I_R\\
 M_2 & M''_{T} & M_T
\end{array}\;\;C \begin{array}{ccc}I_2 & I_3&I_L\\
 M_2 & M''_{T} & M_T 
\end{array}= \delta_{I_L I_R}
$$

\vspace{1cm}
\subsection{$<0s^4\; 0p^2\, |V|\, 0s^4\; 0p^2 >$}

The matrix element can be written in obvious notation as
$$V_{ss}+V_{pp}+V_{sp}\;.$$
The first term is just $<0s^4\; |V_{ss}|\, 0s^4\; >$ and the 
second is $< 0p^2\, |V_{pp}|\, 0p^2 >$.  Only the third term 
is new.

\begin{eqnarray}
\hspace{-1.9cm}&&\Big< (0s)^4\,  [f_4]_{c}(\lambda_4\mu_4) \;S_4 \;I_4  \times  
(0p)^{2}\,[f_{2}]_{c}(\lambda_{2}\mu_{2})_{c}\,L_{2}\,S_{2}I_{2}\;;(0 0)_c
\;L\;S\;JI\Big|\;\sum_{i,j}V_{ij}\;\Big|\nonumber\\
&&\hspace{5.5cm}\Big|(0s)^4\, \, [\bar{f_4}]_{c} 
(\bar{\lambda_4}\bar{\mu_4})\;\bar{S_4} \;\bar{I_4} 
\times (0p)^{2}\;[\bar{f_{2}}]_{c}(\bar{\lambda_{2}}\bar{\mu_{2}})_{c}\,\bar{L_{2}}\
,\bar{S_{2}}\bar{I_{2}}\;;(0 0)_c
\;\bar{L}\;\bar{S}\;J\,I\Big>\; \nonumber\\ \nonumber\\
&&=\;\Big< (0s)^4\,  [f_4]_{c} (\lambda_4\mu_4)\;S_4 \;I_4\Big|\;\sum_{i,j}V_{ij}\;\Big|
(0s)^4\, \, [\bar{f_4}]_{c} (\bar{\lambda_4}\bar{\mu_4})\;\bar{S_4} \;\bar{I_4} \Big>\;\delta_{I_{2} \bar{I_{2}}} 
\;\delta_{S_{2} \bar{S_{2}}}\;\delta_{2 \bar{2}}
\;+\nonumber\\
&&+\Big<(0p)^{2}\,[f_{2}]_{c}(\lambda_{2}\mu_{2})_{c}\,L_{2}\,S_{2}\Big|\;\sum_{{i<j}}V\;\Big|
(0p)^{2}\,[\bar{f_{2}}]_{c}(\bar{\lambda_{2}}\bar{\mu_{2}})_{c}\,L_{2}\,S_{2}\Big>
\delta_{1 \bar{1}}\nonumber\\
&&+\Big< (0s)^4\,  [f_4]_{c} (\lambda_4\mu_4)\;S_4 \;I_4  \times  
(0p)^{2}\,[f_{2}]_{c}(\lambda_{2}\mu_{2})_{c}\,L_{2}\,S_{2}I_{2}\;;(0 0)_c
\;L\;S\;JI\Big|\;\sum_{i,j}V_{ij}\;\Big|\nonumber\\
&&\hspace{5.5cm}\Big|(0s)^4\, \, 
[\bar{f_4}]_{c}(\bar{\lambda_4}\bar{\mu_4}) \;\bar{S_4} \;\bar{I_4} 
\times (0p)^{2}\;[\bar{f_{2}}]_{c}(\bar{\lambda_{2}}\bar{\mu_{2}})_{c}\,\bar{L_{2}}\
,\bar{S_{2}}\bar{I_{2}}\;;(0 0)_c
\;\bar{L}\;\bar{S}\;J\,I\Big>\nonumber
\end{eqnarray}
%\newpage
This last matrix element can be evaluated using
\begin{eqnarray}
& &\Big< (0s)^4\,  [f_4]_{c} (\lambda_4\mu_4)\;S_4 \;I_4  \times  
(0p)^{2}\,[f_{2}]_{c}(\lambda_{2}\mu_{2})_{c}\,L_{2}\,S_{2}I_{2}\;;(0 0)_c
\;L\;S\;JI\Big|\;\sum_{i,j}V_{ij}\;\Big|\nonumber\\
&&\hspace{5.5cm}\Big|(0s)^4\, \, 
[\bar{f_4}]_{c}(\bar{\lambda_4}\bar{\mu_4}) \;\bar{S_4} \;\bar{I_4} 
\times (0p)^{2}\;[\bar{f_{2}}]_{c}(\bar{\lambda_{2}}\bar{\mu_{2}})_{c}\,\bar{L_{2}}\
,\bar{S_{2}}\bar{I_{2}}\;;(0 0)_c
\;\bar{L}\;\bar{S}\;J\,I\Big>\nonumber\\
&=& 8\sum \sqrt{\frac{n_{f'_3} n_{f_3}}{n_{f_4}n_{\bar{f}_4}}} 
    \Bigg\langle\begin{array}{cc}[f_3']_{c}  & [1]_{c}\\
(\lambda_3' \mu_3')_c \,S_3' &(1 0)_c\,\frac{1}{2}\\ 
\end{array}\Bigg|\begin{array}{c}[f_4]_{c}\\ (\lambda_4\mu_4) \;S_4 
\end{array}
\Bigg\rangle
\Bigg\langle\begin{array}{cc}[\bar{f}_3]_{c}  & [1]_{c}\\
(\bar{\lambda_3} \bar{\mu}_3)_c \,\bar{S}_3 &(1 0)_c\,\frac{1}{2}\\ 
\end{array}\Bigg|\begin{array}{c}[\bar{f}_4]_{c}\\ (\bar{\lambda}_4\bar{\mu}_4) 
\;\bar{S}_4 
\end{array}
\Bigg\rangle\nonumber\\
 &&\times \;  \Bigg\langle\begin{array}{cc}[f_{2}']_{c}  & [1]_{c}\\
(1 0)_c \,\frac{1}{2} &(1 0)_c\,\frac{1}{2}\\ 
\end{array}\Bigg|\begin{array}{c}[f_{2}]_{c}\\ (\lambda_{2}\mu_{2}) \;S_2 
\end{array}
\Bigg\rangle\Bigg\langle\begin{array}{cc}[f_{2}']_{c}  & [1]_{c}\\
(1 0)_c \,\frac{1}{2} &(1 0)_c\,\frac{1}{2}\\ 
\end{array}\Bigg|\begin{array}{c}[\bar{f}_{2}]_{c}\\ 
(\bar{\lambda_{2}}\bar{\mu_{2}}) \;\bar{S}_2 
\end{array}
\Bigg\rangle\nonumber\\
&&\times \sum    U \Bigg(\begin{array}{ccc}0 &  0 &0\\
1 & 1& L_{2}\\ 1 & 1 & L 
\end{array}
\Bigg)
  U \Bigg(\begin{array}{ccc}0 &  0&0\\
1 & 1& \bar{L_{2}}\\ 1 &1 & \bar{L} 
\end{array}
\Bigg)
U \Bigg(\begin{array}{ccc}S_3' &  \frac{1}{2} &S_4\\
\frac{1}{2} & \frac{1}{2}& S_{2}\\ S' & S_{12} & S 
\end{array}
\Bigg)   %---------------
  U \Bigg(\begin{array}{ccc}S_3' &  \frac{1}{2}&\bar{S_4}\\
\frac{1}{2} & \frac{1}{2}& \bar{S_{2}}\\ S' & \bar{S}_{12} & \bar{S} 
\end{array}
\Bigg)\nonumber\\
&&\times \;\; \sum_{J' J_{12}}  U \Bigg(\begin{array}{ccc}1 & 1 &L\\
S' & S_{12}& S\\ J' & J_{12} & J 
\end{array}
\Bigg)
U \Bigg(\begin{array}{ccc}1 &1&\bar{L}\\
S' & \bar{S}_{12}& \bar{S}\\ J' & J_{12} & J 
\end{array}
\Bigg)\nonumber\\
\times&&
U\left(\begin{array}{ccc}(\lambda_3' \mu_3')_c &(1 0)&(\lambda_4 \mu_4)_c\\
(1 \,0) & (1\, 0)& (\lambda_2 \mu_2)_c\\ (\lambda' \mu')_c & (\lambda_{12} \mu_{12})_c & (0\, 0) 
\end{array}\right)
U\left(\begin{array}{ccc}(\lambda_3' \mu_3')_c &(1 0)&(\lambda_4 \mu_4)_c\\
(1 0) & (1 0)& (\lambda_2 \mu_2)_c\\ (\lambda' \mu')_c & (\lambda_{12} \mu_{12})_c & (0 \,0) 
\end{array}\right)
\nonumber\\
\times&&
\Big<0s\,0p\;L_{12}S_{12} J_{12} \Big| V |0s\,0p\;L_{12}\bar{S}_{12} J_{12}\Big>\label{eq:s4p2}
\end{eqnarray}

One can now simplify; the two orbital angular momentum nine-Js are 
unity.  
Also,

\begin{eqnarray}
&&\hspace{3cm}U\left(\begin{array}{ccc}(\lambda_1 \mu_1)_c &(\lambda_2 \mu_2)_c&(\lambda_{12} \mu_{12})_c\\
(\lambda_3 \mu_3)_c & (\lambda_4 \mu_4)_c& (\mu_{12}\lambda_{12} )_c\\ (\lambda_{13} \mu_{13})_c & (\mu_{13}\lambda_{13} )_c & (0\, 0) _c
\end{array}\right)\nonumber\\
&&=(-)^{\lambda_{1}+\mu_{1}+\lambda_{13}+\mu_{13}-\lambda_{12}-\mu_{12}
-\lambda_{4}-\mu_{4}}
U\left((\lambda_2 \mu_2)_c(\lambda_1 \mu_1)_c (\lambda_4 \mu_4)_c(\lambda_3 \mu_3)_c;(\lambda_{12} \mu_{12})_c(\lambda_{34} \mu_{34})_c\right)\end{eqnarray}
which in our case becomes
$$
(-)^{\lambda_{3}'+\mu_{3}'+\lambda'+\mu'-1
-\lambda_{4}-\mu_{4}}
U\left((1 0)_c(\lambda_3' \mu_3')_c (1 0)_c(1 0)_c;(\lambda_{4} \mu_{4})_c (\mu_{4} \lambda_{4})_c\right)\,\;\;\;\;{\it check}
$$

\hspace{1.cm}

%\subsection{2 $\hbar\omega$ Excitations}
\hspace{.5cm}

\subsection{$<0s^6 \,|V| \, 0s^{4}\; 0p^2>$}
\begin{eqnarray}
&&\hspace{-.9cm}< (0s)^6 \; [f]_{c}\;(0 0)_c \;S
\;I\;\Big|\;\sum_{i,j}V_{ij}\;\Big|\;
(0s)^4\,  [f_{4}]_{cs}(\lambda_{4}\mu_{4}) \;S_4 \;I_4  \times  
(0p)^{2}\,[f_{2}]_{cs}(\lambda_{2}\mu_{2})_{c}\,L_{2}\,S_{2}I_{2}\;;(0 0)_{cs}
\;L\;S\;JI>
\nonumber\\
&&= \sum_{f", f'_{12} S'_{12} \cdots} \;\;\sqrt{\frac{n_{f''}}{n_{f}}}\;\Bigg\langle\begin{array}{cc}[f'']_{cs}  & [f_{12}]_{cs}\\
(\lambda'' \mu'')_{cs} \,S'' &(\mu'' \lambda'')_{cs}\,S_{12}\\ 
\end{array}\Bigg|\begin{array}{c}[f]_{cs}\\(0 0)_{cs} \;S 
\end{array}
\Bigg\rangle
\nonumber\; \;U \left(
\begin{array}{ccc}
0&0&0\\
S_4&S_{12}&S\\S_4&J_{12}&J
\end{array}\right)_L\;
U \left(
\begin{array}{ccc}
0&L_2&L\\
S_4&S_{2}&S\\S_4&J_{12}&J
\end{array}\right)_R
\nonumber\\
&&\times
< (0s)^2 \; [f_{12}]_{cs}\;(\mu'' \lambda'')_{cs}\;S_{12}
\;J_{12}\,I_2\;\Big|\;V_{12}\;\Big|\; 
(0p)^{2}\,[f_{2}]_{cs}(\lambda_{2}\mu_{2})_{c}\,L_{2}\,S_{2}\,J_{12}\,I_{2}\;>
\nonumber
\\  %____________________
&&= (n factor) \sum_{ f_{12}, S_{12},S_4 \cdots} \;\;\sqrt{\frac{n_{f_4}}{n_{f}}}\;\Bigg\langle\begin{array}{cc}[f_4]_{cs}  & [f_{12}]_{cs}\\
(\lambda_4 \mu_4)_{cs} \,S_4 &(\mu_4 \lambda_4)_{cs}\;S_{12}\\ 
\end{array}\Bigg|\begin{array}{c}[f]_{cs}\\(0 0)_{cs} \;S 
\end{array}
\Bigg\rangle \;\;\;U \left(
\begin{array}{ccc}
0&L_2&L\\
S_4&S_{2}&S\\S_4&J_{12}&J
\end{array}\right)_R
\nonumber\; \;
\\
&&\times
< (0s)^2 \; [f_{12}]_{cs}\;(\mu_4 \lambda_4)_{cs}\;S_{12}
\;J_{12}\,I_2\;\Big|\;V_{12}\;\Big|\; 
(0p)^{2}\,[f_{2}]_{cs}(\lambda_{2}\mu_{2})_{c}\,L_{2}\,S_{2}\,J_{12}\,I_{2}\;>
\end{eqnarray}

\hspace{1cm}

\subsection{$<0s^6 \,|V| \, 0s^{5}\; 1s0d >$}

\begin{eqnarray}
&&\hspace{-.9cm}< (0s)^6 \; [f_6]_{cs}\;(0 0)_c \;S_6
\;I\;\Big|\;\sum_{i,j}V_{ij}\;\Big|\;
(0s)^5\,  [f_{5}]_{cs}(0 \,1) \;S_{5} \;I_{5}  \times  
(1s0d)\;\,[1]_{cs}(1 0)_{cs}\;L_{2}\,\frac{1}{2}I_{2}\;;(0 0)_{cs}
\;L\;S_R\;JI>
\nonumber\\
&&= \sum_{f", f'_{12} S''_{12} \cdots} \;\;\Bigg\langle\begin{array}{cc}[f'']_{cs}  & [f_{12}]_{cs}\\
(\lambda'' \mu'')_{cs} \,S'' &(\mu'' \lambda'')_{c}\,S''_{12}\\ 
\end{array}\Bigg|\begin{array}{c}[f_6]_{c}\\(0 0)_{c} \;S_6 
\end{array}
\Bigg\rangle
\nonumber\; U \left(
\begin{array}{ccc}
0&0&0\\
S''&S_{12}''&S_6\\J''&J_{12}&J
\end{array}\right)\;}\sqrt{\frac{n_{f''}}{n_{f}}\;
\nonumber\\
&&\sum_{f'',(\lambda'',\mu''),S',I',[f_2],I_2}\sqrt{\frac{n_{f''}}{n_{f_{5}}}}\;
    \Bigg\langle\begin{array}{cc}[f'']_{cs}  & [1]_{cs}\\
(\lambda'' \mu'')_c \,S'' &(1 0)_c\,\frac{1}{2}\\ 
\end{array}\Bigg|\begin{array}{c}[f_{5}]_{cs}\\(0 1)_c \;S_{5} 
\end{array}
\Bigg\rangle\nonumber\\
&\times&
< \Bigg[(0s)^4\, \, [f'']_{cs}\,(\lambda''\mu'')\;S'' \;I''\times \,(0s)^2 \; [f_{12}]_{cs}\;(\mu'' \lambda'')_{cs}\;S''_{12}
\Bigg]^{S_6JI}\;\Big|\;V_{12}\;\Big|
\nonumber\\&&
\Big|\Bigg[(0s)^4\, \, [f'']_{cs}\,(\lambda''\mu'')\;S'' \;I''\times 0s\,
[1]_{cs}\, (1 0)_{c} 
\;\frac{1}{2}\,\frac{1}{2}\Bigg]^{{S_{5}\,(01)_{c}}}\times\;
1s0d\;(1 0)_{c}\,l\;\frac{1}{2}\,\frac{1}{2}\,;(0 0)_c
\;L\;S_R\;J>\nonumber\\
&&= \sum_{f", f_{12} S''_{12} \cdots} \;\;\Bigg\langle\begin{array}{cc}[f'']_{cs}  & [f_{12}]_{cs}\\
(\lambda'' \mu'')_{cs} \,S'' &(\mu'' \lambda'')_{cs}\,S''_{12}\\ 
\end{array}\Bigg|\begin{array}{c}[f_6]_{cs}\\(0 0)_{cs} \;S_6
\end{array}
\Bigg\rangle 
\nonumber\; \;
\nonumber\\
&&\sum_{f'',(\lambda'',\mu''),S'',I'',[f_2],I_2}\sqrt{\frac{n_{f''}^2}{n_{f_5}n_{f_6}}}\;
    \Bigg\langle\begin{array}{cc}[f'']_{cs}  & [1]_{cs}\\
(\lambda'' \mu'')_c \,S'' &(1 0)_c\,\frac{1}{2}\\ 
\end{array}\Bigg|\begin{array}{c}[f_{5}]_{cs}\\(0 1)_c \;S_{5} 
\end{array}
\Bigg\rangle \nonumber\\
&\times&
\hspace{-1.2cm}\sum_{(\lambda_{12}\mu_{12})S_{12}I_{12}J_{12}J''} 
U\bigg(  (\lambda'' \mu'') (1 0) (0  \,0) (1\,0) ; (0\,1) ( \mu'' \lambda'')\bigg)\; U\Big(S''\frac{1}{2}S_R\frac{1}{2};S_{5}S_{12}\Big)
U\Big(I''\frac{1}{2}I\frac{1}{2};I_{5}I_{12}\Big)\,
\nonumber\\
&&\times \,U\left( \begin{array}{ccc}0 & \ell&\ell\\
S''&S''_{12}&S_R\\S''&J_{12}&J
\end{array}\right)\nonumber
\\
&\times&< \Bigg[(0s)^4\, \, [f'']_{cs}\,(\lambda''\mu'')\;S''\;I''\times \,(0s)^2 \; [f_{12}]_{cs}\;(\mu'' \lambda'')_{cs}\;S_{12}
\;J_{12}\,I_{12}\Bigg]^{S_6JI}\;\Big|\;V_{12}\;\Big|
\nonumber\\
&& \Bigg|\,(0s)^4\, \, [f']_{cs}\,(\lambda''\mu'')\;S'' \;I'' J'\times \Bigg[0s\,
[1]_{cs}\, (1 0)_{c} 
\;\frac{1}{2}\,\frac{1}{2}\;\times
1s0d\;(1 0)_{c}\,l\;\frac{1}{2}\,\frac{1}{2}\Bigg]^{(\lambda_{12}\mu_{12})L_{12}{S_{12}I_{12}\,J_{12}	}}\,;(0 0)_c
\;J>\nonumber
\end{eqnarray}

The last matrix element becomes
$$
<  \,(0s)^2 \; [f_{12}]_{cs}\;(\mu'' \lambda'')_{cs}\;S''_{12}
\;J_{12}\,I_{12}\;\Big|\;V_{12}\;\Big| \;0s\,
[1]_{cs}\, (1 0)_{c} 
\;\;\times
1s0d\;(1 0)_{c}\,l\;\,;(\lambda_{12}\mu_{12})\,L_{12}\,S_{12}\,J_{12}I_{12}>
$$
$$
\equiv < (0s)^2 \; \;(\mu'' \lambda'')_{cs}\;S''_{12}
\;J_{12}\,I_{12}\;\Big|\;V_{12}\;\Big|
\;0s\, \times
1s0d\;;(\mu''\lambda'')\,L_{12}\,S_{12}\,J_{12}I_{12}>
$$
Note that the two-body matrix element is diagonal in SU(3) by virtue that both the $0s^2$ particles and the $0s \,0s1d$ pair must couple to the $(\lambda''\mu'')$ states to form a color singlet.
Finally, one has
\begin{eqnarray}
&&\hspace{-2cm}\sum_{f", f_{12} ,(\lambda''\mu''),S'',I'',S_{12},I_{12} \cdots} \;\;\hspace{-.4cm}f(N)\Bigg\langle\begin{array}{cc}[f'']_{cs}  & [f_{12}]_{cs}\\
(\lambda'' \mu'')_{cs} \,S'' &(\mu'' \lambda'')_{cs}\,S'_{12}\\ 
\end{array}\Bigg|\begin{array}{c}[f_6]_{cs}\\(0 0)_{cs} \;S_6
\end{array}
\Bigg\rangle 
\sqrt{\frac{n_{f''}^2}{n_{f_5}n_{f_6}}}\;
    \Bigg\langle\begin{array}{cc}[f'']_{cs}  & [1]_{cs}\\
(\lambda'' \mu'')_c \,S'' &(1 0)_c\,\frac{1}{2}\\ 
\end{array}\Bigg|\begin{array}{c}[f_{5}]_{cs}\\(0 1)_c \;S_{5} 
\end{array}
\Bigg\rangle
\nonumber\\
&\times&
\hspace{-.2cm} 
U\bigg(  (\lambda'' \mu'') (1 0) (0  \,0) (1\,0) ; (0\,1) ( \mu'' \lambda'')\bigg)\; U\Big(S''\frac{1}{2}S_R\frac{1}{2};S_{5}S_{12}\Big)\;\;
U\Big(I''\frac{1}{2}I\frac{1}{2};I_{5}I_{12}\Big)\,
\,U\left( \begin{array}{ccc}0 & \ell&\ell\\
S''&S''_{12}&S_R\\S''&J_{12}&J
\end{array}\right)\nonumber
\\
&\times&< (0s)^2 \; \;(\mu'' \lambda'')_{cs}\;S''_{12}
\;J_{12}\,I_{12}\;\Big|\;V_{12}\;\Big|
\;0s\, \times
1s0d\;\ell\;;(\mu'' \lambda'')\,L_{12}\,S_{12}\,J_{12}I_{12}>\nonumber
\end{eqnarray}
\\\\\\\\

\subsubsection{Example}

\begin{eqnarray}
&&\hspace{-.9cm}< (0s)^6 \; [2211]_{cs}\;(0 0)_c \;S_6=0
\;I=1\;\Big|\;\sum_{i,j}V_{ij}\;\Big|\;\nonumber\\
&&\hspace{.1cm}\bigg|
(0s)^5\,  [221]_{cs}(0 \,1) \;S_{5}=\frac{1}{2} \;I_{5}=\frac{1}{2}  \times  
(1s0d)\;\,[1]_{cs}(1 0)_{c}\;L_{2}=0\,\frac{1}{2}I_{2}=\frac{1}{2}\;;(0 0)_{c}
\;L=0\;S_R=0\;\;J=0\;I=1>\nonumber\\
&&\hspace{-2cm}=\hspace{-1cm}\sum_{f", f_{12} ,(\lambda''\mu''),I'',S_{12},I_{12} \cdots} \;\;\hspace{-1.3cm}f(N)\;\Bigg\langle\begin{array}{cc}[f'']_{cs}  & [f_{12}]_{cs}\\
(\lambda'' \mu'')_{c} \,S'_{12} &(\mu'' \lambda'')_{c}\,S'_{12}\\ 
\end{array}\Bigg|\begin{array}{c}[2211]_{cs}\\(0 0)_{cs} \;S_6=0
\end{array}
\Bigg\rangle 
\sqrt{\frac{n_{f''}^2}{5\cdot 9}}\;
    \Bigg\langle\begin{array}{cc}[f'']_{cs}  & [1]_{cs}\\
(\lambda'' \mu'')_c \,S'_{12} &(1 0)_c\,\frac{1}{2}\\ 
\end{array}\Bigg|\begin{array}{c}[221]_{cs}\\(0 1)_c \;S_{5} =\frac{1}{2}
\end{array}
\Bigg\rangle
\nonumber\\
&\times&
\hspace{-.2cm} 
U\bigg(  (\lambda'' \mu'') (1 0) (0  \,0) (1\,0) ; (0\,1) ( \mu'' \lambda'')\bigg)\; U\Big(S''\,\frac{1}{2}\,0\frac{1}{2};\frac{1}{2}\,S_{12}\Big)\;\;
U\Big(I''\;\frac{1}{2}\;1\;\frac{1}{2};\frac{1}{2}\;I_{12}\Big)\,
\,U\left( \begin{array}{ccc}0 & 0&0\\
S''&S'_{12}&0\\S''&J_{12}&J=0
\end{array}\right)\nonumber
\\
&\times&< (0s)^2 \; \;(\mu'' \lambda'')_{cs}\;S'_{12}
\;J_{12}\,I_{12}\;\Big|\;V_{12}\;\Big|
\;0s\, \times
1s\;;(\mu'' \lambda'')\,L_{12}=0\,S_{12}\;J_{12}I_{12}>\nonumber
\end{eqnarray}
\hspace{2cm}
%\newpage

\subsection{$<0s^4\; 0p^2\, |V|\, 0s^5 \; 1s0d  >$}

One can use the results from above, namely Eq.~(\ref{eq:s5sd}) and Eq. (\ref{eq:s4p2}), to obtain

\begin{eqnarray}
&&\Big< (0s)^4\,  [f_4]_{cs}(\lambda_4\mu_4) \;S_4 \;I_4  \times  
(0p)^{2}\,[f_{2}]_{cs}(\lambda_{2}\mu_{2})_{c}\,L_{2}\,S_{2}I_{2}\;;(0 0)_c
\;L\;S\;JI\Big|\;\sum_{i,j}V_{ij}\;\Big| 
\nonumber\\
&&\hspace{7cm}\times\;\Big|\; (0s)^5\, \, [f]_{cs} \;S_5 \;I_5 \times 
1s0d\;\;;(0 1)_c
\;L_R\;S_R\;J\;\Big>\nonumber\\
&=&\hspace{-.2cm}\sum_{(\lambda_{12}\mu_{12})S_{12}I_{12}J_{12}} \hspace{-.42cm}
U\Big( (\lambda_4\mu_4) (1  0) (0 0) (1 0); (0 1) 
(\lambda_{12}\mu_{12})\Big)U\Big(S_4\frac{1}{2}S_R\frac{1}{2};S_5S_{12}\Big)
U\Big(I_4\frac{1}{2}I_R\frac{1}{2};I_5I_{12}\Big)\nonumber
\\
&&\times\hspace{1cm}\sqrt{\frac{n_{f_4}}{n_{f_{5}}}}\;
    \Bigg\langle\begin{array}{cc}[f_4]_{cs}  & [1]_{cs}\\
(\lambda_4 \mu_4)_c \,S_4 &(1 0)_c\,\frac{1}{2}\\ 
\end{array}\Bigg|\begin{array}{c}[f_{5}]_{cs}\\(0 1)_c \;S_{5} 
\end{array}
\Bigg\rangle
U
    \Bigg(\begin{array}{ccc}0 & L_{2} &L\\
S_4 & S_{2}& S\\ J_4 & J_{12} & J
\end{array}\Bigg)_L
  U
    \Bigg(\begin{array}{ccc}0 & L_{12} &L_R\\
S_4 & S_{12}& S_R\\ J_4 & J_{12} & J
\end{array}\Bigg)_R
\nonumber\\
&\times&\Big< (0s)^4\,  [f_4]_{cs}(\lambda_4\mu_4) \;S_4 \;I_4  \times  
(0p)^{2}\,[f_{2}]_{cs}(\lambda_{2}\mu_{2})_{c}\,L_{2}\,S_{2}I_{2}\;;(0 0)_c
\;L\;S\;JI\Big|\;\sum_{i,j}V_{ij}\Big|\;
\nonumber 
\\
&\times&\hspace{0.2cm}\Big|\,(0s)^4\, \, [f_4]_{cs}\,(\lambda_4\mu_4)\;S_4 \;I_4 J_4
\times \Bigg[0s\,
[1]_{cs}\, (1 0)_{c} 
\;\frac{1}{2}\,\frac{1}{2}\;\times
1s0d\;(1 0)_{c}\,l\;\frac{1}{2}\,\frac{1}{2}\Bigg]^{(\lambda_{12}\mu_{12})L_{12}{S_{12}I_{12}\,J_{12}}}\,;(0 0)_c
\;J\;\;\Big>
\nonumber\\
&=&\hspace{-.42cm}\sum_{S_{12}J_{12}} \hspace{.12cm}
U\Big( (\lambda_4\mu_4) (1  0) (0 0) (1 0); (0 1) 
(\lambda_{12}\mu_{12})\Big)U\Big(S_4\frac{1}{2}S_R\frac{1}{2};S_5S_{12}\Big)
U\Big(I_4\frac{1}{2}I_R\frac{1}{2};I_5I_{12}\Big)\nonumber
\\
&&\times\hspace{1cm}\sqrt{\frac{n_{f_4}}{n_{f_{5}}}}\;
    \Bigg\langle\begin{array}{cc}[f_4]_{cs}  & [1]_{cs}\\
(\lambda_4 \mu_4)_c \,S_4 &(1 0)_c\,\frac{1}{2}\\ 
\end{array}\Bigg|\begin{array}{c}[f_{5}]_{cs}\\(0 1)_c \;S_{5} 
\end{array}
\Bigg\rangle \;\; U
    \Bigg(\begin{array}{ccc}0 & L_{2} &L\\
S_4 & S_{2}& S\\ J_4 & J_{12} & J
\end{array}\Bigg)_L    
U
    \Bigg(\begin{array}{ccc}0 & L_{12} &L_R\\
S_4 & S_{12}& S_R\\ J_4 & J_{12} & J
\end{array}\Bigg)_R
\nonumber\\
&\times&\Big<  
(0p)^{2}\,[f_{2}]_{cs}(\lambda_{2}\mu_{2})_{c}\,L_{2}\,S_{2}I_{2}\;J_{12}
\Big|\;V_{ij}\Big|\;
0s\,\, (1 0)_{c} 
\;\times\;
1s0d\;(1 0)_{c}\,l\,  ;   (\lambda_{12}\mu_{12})\;S_{12}I_{12}\,J_{12}	
\;\Big>
\end{eqnarray}
One could simplify the $9j$ and the $SU(3)$ Racah coefficient, but it is preferable to leave it in the above form and let the code fix the phases.
There is no sum over $(\lambda' \mu')$ because of the SU(3) Racah and the coupling to $(0 0)$.
\newpage
\subsubsection{Example 1}

Assume $J=I=0$.
\begin{eqnarray}
 &&< [211]\; (1 0)_c \;S_4=1 \;I_4=1 \times [11]\, (0 1)_c L_2=0\,S_2=1\; I_2=1; \,L=0 \,S=0 | V | \nonumber\\
&&\hspace{2cm}\times [221]\; (0 1)_c \;S_5=\frac{1}{2} \,I_5=\frac{1}{2} \times\, [1] \frac{1}{2}\frac{1}{2}\,\ell= 0 \,L=0 S=0>\nonumber\\
&=&\hspace{-.2cm}\sum_{S_{12}} \hspace{.32cm}
U\Big( (1 0) (1  0) (0 0) (1 0); (0 1) 
(0 1)\Big)U\Big(1\frac{1}{2}0\frac{1}{2};\frac{1}{2} S_{12}\Big)
U\Big(1\frac{1}{2}0\frac{1}{2};\frac{1}{2}1\Big)\nonumber
\\
&&\times\hspace{1cm}\sqrt{\frac{3}{5}}\;
    \Bigg\langle\begin{array}{cc}[211]_{cs}  & [1]_{cs}\\
(1 0)_c \,1 &(1 0)_c\,\frac{1}{2}\\ 
\end{array}\Bigg|\begin{array}{c}[221]_{cs}\\(0 1)_c \;\frac{1}{2} 
\end{array}
\Bigg\rangle \;\; U
    \Bigg(\begin{array}{ccc}0 & \ell=0 &0\\
1 & S_{12}& 0\\ 1 & S_{12} & 0
\end{array}\Bigg)
\nonumber\\
&\times&\Big<  
(0p)^{2}\,[11]_{cs}(0 1)_{c}\,L_{2}=0\,S_{2}=1I_{2}=1\;J_{2}=1
\Big|\;V_{ij}\Big|\;
0s\,\,  
\;\times\;
1s0d\;\,\ell=0\,  ;   (0 1)\;S'_{12}I_{12}=1\,J_{12}	
\;\Big>
\nonumber
\end{eqnarray}

We have:
\begin{eqnarray}
\sqrt{\frac{n_{f'}}{n_f}}&=&\sqrt{\frac{3}{5}}\nonumber\\
 \Bigg\langle\begin{array}{cc}[211]_{cs}  & [1]_{cs}\\
(1 0)_c \,1 &(1 0)_c\,\frac{1}{2}\\ 
\end{array}\Bigg|\begin{array}{c}[221]_{cs}\\(0 1)_c \;\frac{1}{2} 
\end{array}
\Bigg\rangle  &=&  -\sqrt{\frac{2}{3}}\nonumber\\
 \;\; U
    \Bigg(\begin{array}{ccc}0 & \ell=0 &0\\
1 & S_{12}& 0\\ 1 & S_{12} & 0
\end{array}\Bigg)&=&1
\nonumber\\
U\Big( (1 0) (1  0) (0 0) (1 0); (0 1) 
(0 1)\Big)&=&1\nonumber\\
U\Big(1\frac{1}{2}0\frac{1}{2};\frac{1}{2} S_{12}\Big)&=&-1\nonumber\\
U\Big(1\frac{1}{2}0\frac{1}{2};\frac{1}{2}1\Big)&=&-1\nonumber\\
product&=& -\sqrt{\frac{2}{5}}\nonumber
\end{eqnarray}

\newpage
\subsubsection{Example 2}

Assume $J=0, I=1$.
\begin{eqnarray}
 &&< [22]\; (0 2)_c \;S_4=0 \;I_4=0 \times [11]\, (2 0)_c \;L_2=0\,S_2=0\; I_2=1; \,L=0 \,S=0 | V | \nonumber\\
&&\hspace{2cm}\times [221]\; (0 1)_c \;S_5=\frac{1}{2} \,I_5=\frac{1}{2} \times\, [1] \frac{1}{2}\frac{1}{2}\,\ell= 0 \,L=0 S=0>\nonumber\\
&=&\hspace{-.2cm}\sum_{S_{12}} \hspace{.32cm}
U\Big( (0 2) (1  0) (0 0) (1 0); (0 1) 
(2 0)\Big)U\Big(0\frac{1}{2}0\frac{1}{2};\frac{1}{2} S_{12}=0\Big)
U\Big(0\frac{1}{2}0\frac{1}{2};\frac{1}{2}0\Big)\nonumber
\\
&&\times\hspace{1cm}\sqrt{\frac{3}{5}}\;
    \Bigg\langle\begin{array}{cc}[22]_{cs}  & [1]_{cs}\\
(0 2)_c \,1 &(1 0)_c\,\frac{1}{2}\\ 
\end{array}\Bigg|\begin{array}{c}[221]_{cs}\\(0 1)_c \;\frac{1}{2} 
\end{array}
\Bigg\rangle \;\; U
    \Bigg(\begin{array}{ccc}0 & \ell=0 &0\\
1 & S_{12}=0& 0\\ 1 & S_{12} & 0
\end{array}\Bigg)
\nonumber\\
&\times&\Big<  
(0p)^{2}\,[11]_{cs}(0 1)_{c}\,L_{2}=0\,S_{2}=1I_{2}=1\;J_{2}=1
\Big|\;V_{ij}\Big|\;
0s\,\,  
\;\times\;
1s0d\;\,\ell=0\,  ;   (0 1)\;S'_{12}I_{12}=1\,J_{12}	
\;\Big>
\nonumber
\end{eqnarray}

one has:
\begin{eqnarray}
\sqrt{\frac{n_{f'}}{n_f}}&=&\sqrt{\frac{2}{5}}\nonumber\\
 \Bigg\langle\begin{array}{cc}[211]_{cs}  & [1]_{cs}\\
(1 0)_c \,1 &(1 0)_c\,\frac{1}{2}\\ 
\end{array}\Bigg|\begin{array}{c}[221]_{cs}\\(0 1)_c \;\frac{1}{2} 
\end{array}
\Bigg\rangle  &=&  \sqrt{\frac{3}{4}}\nonumber\\
 \;\; U
    \Bigg(\begin{array}{ccc}0 & \ell=0 &0\\
1 & S_{12}& 0\\ 1 & S_{12} & 0
\end{array}\Bigg)&=&1
\nonumber\\
U\Big( (1 0) (1  0) (0 0) (1 0); (0 1) 
(0 1)\Big)&=&1\nonumber\\
U\Big(1\frac{1}{2}0\frac{1}{2};\frac{1}{2} S_{12}\Big)&=&1\nonumber\\
U\Big(1\frac{1}{2}0\frac{1}{2};\frac{1}{2}1\Big)&=&-1\nonumber\\
product&=& \sqrt{\frac{3}{10}}\nonumber
\end{eqnarray}
This multiplies the 2BME
$$\langle\, 0p^2 [11]_{cs} (2 0)_c S_2=0\;I_2=1\,J_2=1\bigg|V\bigg| 0s\times 1s\,
(2 0)_c S_{12}=0 \,J_{12}=1 \,I_{12}=1\, \rangle.$$

\subsubsection{Example 3}

Assume $J=4, I=1$.
\begin{eqnarray}
 &&< [22]\; (0 2)_c \;S_4=2 \;I_4=0 \times [11]\, (2 0)_c \;L_2=2\,S_2=0\; I_2=1; \,\;L=2 \,\;S=2 \;\;J=4 \,| V | \nonumber\\
&&\hspace{2cm}\times [221]\; (0 1)_c \;S_5=\frac{5}{2} \;I_5=\frac{1}{2} \times\, [1] \frac{1}{2}\frac{1}{2}\,\ell= 2 \,L=2 S=2\;\;J=4\;I=1/>\nonumber\\
&=&\hspace{-.2cm}\sum_{S_{12}} \hspace{.32cm}
U\Big( (0 2) (1  0) (0 0) (1 0); (0 1) 
(2 0)\Big)U\Big(2\frac{1}{2}2\frac{1}{2};\frac{5}{2} S_{12}\Big)
U\Big(0\frac{1}{2}1\frac{1}{2};\frac{1}{2}I_{12}\Big)\nonumber
\\
&&\times\hspace{1cm}\sqrt{\frac{3}{5}}\;
    \Bigg\langle\begin{array}{cc}[22]_{cs}  & [1]_{cs}\\
(0 2)_c \,1 &(1 0)_c\,\frac{1}{2}\\ 
\end{array}\Bigg|\begin{array}{c}[221]_{cs}\\(0 1)_c \;\frac{1}{2} 
\end{array}
\Bigg\rangle \;\; U
    \Bigg(\begin{array}{ccc}0 & \ell=0 &0\\
1 & S_{12}=0& 0\\ 1 & S_{12} & 0
\end{array}\Bigg)
\nonumber\\
&\times&\Big<  
(0p)^{2}\,[11]_{cs}(0 1)_{c}\,L_{2}=0\,S_{2}=1I_{2}=1\;J_{2}=1
\Big|\;V_{ij}\Big|\;
0s\,\,  
\;\times\;
1s0d\;\,\ell=0\,  ;   (0 1)\;S'_{12}I_{12}=1\,J_{12}	
\;\Big>
\nonumber
\end{eqnarray}

%\end{document}

\newpage
\section{Appendix}

\subsection{SU(2) Conventions}

Use the de-Shalit-Talmi \cite{Talmi} conventions for reduced matrix elements:
\begin{eqnarray}
\big< J M \Big|\, T^{(k)}_{(\kappa)}\, \Big| J' \,M' \big> &=& 
(-)^{J-M}\Big(
\begin{array}{ccc}J& k& J'\\-M & \kappa & M' \end{array}
\Big) \big< J \Big|\Big| \,T^{(k)}\,\Big|\Big| J' \big>\nonumber\\
\big<\, J\, \Big|\Big| \,1\,\Big|\Big|\, J' \,\big> &=& \hat{J} 
\;\delta_{J\,J'}\nonumber\\\nonumber\\
\big<\, J\, \Big|\Big| \,J \,\Big|\Big|\, J' \,\big> &=& \hat{J} 
\;\delta_{J\,J'}\nonumber
\end{eqnarray}

\subsection{15j Symbols}

\begin{eqnarray}
&&\bigg|  \left[(L_4 \ell_5)^{L_5}\,(S_4 s_5)^{S_5}\right] \times \ell_6 s_6; L S J\;\rangle =\sum_{}\;\; U
    \Bigg(\begin{array}{ccc}L_5 & \ell_6 &L\\
S_5 & s_6& S\\ J_5 & j_{6} & J
\end{array}\Bigg)\bigg|  \left[(L_4 \ell_5)^{L_5}\,(S_4 s_5)^{S_5}\right]^{J_5} \times (\ell_6 s_6)^{j_{6}}  J\;\rangle\nonumber\\
&&=\sum_{}\;\; U
    \Bigg(\begin{array}{ccc}L_5 & \ell_6 &L\\
S_5 & s_{6}& S\\ J_5 & j_{6} & J
\end{array}\Bigg)\;
U
    \Bigg(\begin{array}{ccc}L_4 & \ell_5 &L_5\\
S_4 & s_5& S_5\\ J_4 & j_5 & J_5
\end{array}\Bigg)
\; \bigg|  \left[(L_4 S_4)^{J_4}\,(\ell_5 s_5)^{j_5}\right]^{J_5} \times (\ell_6 s_6)^{j_{6}}  J\;\rangle
\nonumber\\
&&=\sum_{}\;\; U
    \Bigg(\begin{array}{ccc}L_5 & \ell_6 &L\\
S_5 & s_{6}& S\\ J_5 & j_{6} & J
\end{array}\Bigg)\;
U
    \Bigg(\begin{array}{ccc}L_4 & \ell_5 &L_5\\
S_4 & s_5& S_5\\ J_4 & j_5 & J_5
\end{array}\Bigg)
U\bigg(J_4 j_5 J j_6; J_5 J_{56}\bigg)
\; \bigg|  (L_4 S_4)^{J_4}\,\left[(\ell_5 s_5)^{j_5}  (\ell_6 s_6)^{j_{6}}\right]^{J_{56}}  J\;\rangle
\nonumber\\
&&=\sum_{J_5, j_6, J_4, j_5, J_{56}, L_{56}, S_{56}}\;\; U
    \Bigg(\begin{array}{ccc}L_5 & \ell_6 &L\\
S_5 & s_{6}& S\\ J_5 & j_{6} & J
\end{array}\Bigg)\;
U
    \Bigg(\begin{array}{ccc}L_4 & \ell_5 &L_5\\
S_4 & s_5& S_5\\ J_4 & j_5 & J_5
\end{array}\Bigg)
U\bigg(J_4 j_5 J j_6; J_5 J_{56}\bigg)
\; \; U
    \Bigg(\begin{array}{ccc}\ell_5 & \ell_6 &L_{56}\\
s_5 & s_{6}& S_{56}\\ j_5 & j_{6} & J_{56}
\end{array}\Bigg)
\nonumber\\&&\bigg|  (L_4 S_4)^{J_4}\,\left[\ell_5 s_5 \times \ell_6 s_6 ; L_{56} S_{56}\right]^{J_{56}} ; J\;\rangle\label{eq"12j}
\end{eqnarray}
Angular momenta involved: $L_4, \ell_5, \ell_6, L_5, L, S_4, s_5, s_6, S_5, S, j_5, j_6, L_{56}, S_{56}, J$

Alternatively,
\newpage
\begin{eqnarray}
&&\bigg|  \left[(L_4 \ell_5)^{L_5}\,(S_4 s_5)^{S_5}\right] \times \ell_6 s_6; L S J\;\rangle =\bigg|  \left[(L_4 \ell_5)^{L_5}\ell_6\right]^L
\left[ \,(S_4 s_5)^{S_5}s_6\right]^{S}   J\;\rangle\nonumber
\\&&=\sum_{}\;\;
U\bigg(L_4\ell_5L\ell_6;L_5L_{56}\bigg)
U\bigg(S_4 s_5 S s_6;S_5 S_{56}\bigg) \bigg|  \left[L_4 (\ell_5\ell_6)^{L_{56}}\right]^L
\left[ \,S_4 (s_5 s_6)^{S_{56}}\right]^{S}   J\;\rangle\nonumber\\
&&=\hspace{-.5cm}\sum_{L_{56}S_{56}J_4J_{56}}\;\; \hspace{-.5cm}
U\bigg(L_4\ell_5L\ell_6;L_5L_{56}\bigg)
U\bigg(S_4 s_5 S s_6;S_5 S_{56}\bigg) 
U  \Bigg(\begin{array}{ccc}L_4 & L_{56} &L\\
S_4 & S_{56}& S\\ J_4 & J_{56} & J
\end{array}\Bigg)\;
\bigg|  \left[L_4 S_4\right]^{J_4}
\left[ (\ell_5\ell_6)^{L_{56}} (s_5 s_6)^{S_{56}}\right]^{J_{56}}   ;J\;\rangle\nonumber\\
&&=U  \Bigg(\begin{array}{ccccccc}L_4&\ell_5 &L&\ell_6&L_5&L_{56}&\\
S_4 &s_5 &S& s_6&S_5 &S_{56}&\\ J_4 &&&&& J_{56} & J
\end{array}\Bigg)
\bigg|  \left[L_4 S_4\right]^{J_4}
\left[ (\ell_5\ell_6)^{L_{56}} (s_5 s_6)^{S_{56}}\right]^{J_{56}}   ;J\;\rangle\end{eqnarray}

\subsection{SU(3) 9 $\lambda \mu$ and 6 $\lambda \mu$ Special Cases}
\begin{eqnarray}
 &&  U \Bigg(\begin{array}{ccc}(\lambda_{1} \mu_{1}) & (\lambda_{2} 
    \mu_{2}) &(\lambda_{12} \mu_{12} )\\
(\lambda_{3} \mu_{3}) & (\lambda_{4} \mu_{4})& ( \mu_{12}\lambda_{12})\\
(\lambda_{13} \mu_{13}) & ( \mu_{13}\lambda_{13}) & (0\;0)
\end{array}
\Bigg) = \nonumber\\
&&(-)^{{\lambda_{1}+\mu_{1}+\lambda_{13}+\mu_{13}-\lambda_{12}-\mu_{12}
  -\lambda_{4}-\mu_{4}}}\;\; 
  U\Big[(\lambda_{2} \mu_{2})(\lambda_{1} \mu_{1})(\lambda_{4} 
  \mu_{4})(\lambda_{3} \mu_{3});(\lambda_{12} \mu_{12})(\lambda_{34} 
  \mu_{34})\Big]
\end{eqnarray}

\begin{eqnarray}
  U\Big[(\lambda_{2} \mu_{2})(\lambda_{1} \mu_{1})( 
  \mu_{2}\lambda_{2})( \mu_{1}\lambda_{1});( \lambda_{12}\mu_{12})(0\; 
  0)\Big] = (-)^{\lambda_{1}+\mu_{1}+\lambda_{2}+\mu_{2
  }-\lambda_{12}-\mu_{12}}\frac{g(\lambda_{12}\,\mu_{12})}{g(\lambda_{1} \mu_{1})
  g(\lambda_{2} \mu_{2})\;
   }\label{eq:6lmz}
\end{eqnarray}

\begin{eqnarray}
 &&  U \Bigg(\begin{array}{ccc}(\lambda_{1} \mu_{1}) & (\lambda_{2}\, 
    \mu_{2}) &(\lambda_{12}\, \mu_{12})\\
(\lambda_{3} \mu_{3}) & (\lambda_{4} \,\mu_{4})& ( \mu_{12}\,\lambda_{12})\\
(0\;0) & (0\;0) & (0\;0)
\end{array}
\Bigg) = \frac{g(\lambda_{12}\,\mu_{12})}{g(\lambda_{1} \mu_{1})
  g(\lambda_{2} \mu_{2})}
\end{eqnarray}

\begin{eqnarray}
    U\Big( (\lambda'\mu') (1  0) (0 0) (1 0); (0 1) 
(\lambda_{12}\mu_{12})\Big) &=&\delta_{(\lambda'\mu') (1 \,
0)}\delta_{(0\, 1)(\lambda_{12}\mu_{12})}\nonumber
\nonumber\end{eqnarray}

\subsection{Matrix Elements}

\begin{eqnarray}
    \big< J_{1}(1)\, J_{2}(2),\, J \Big| V_{1} \Big| J_{3}(1)\,J_{4}(2),\,J\big> = 
   \delta_{J_{2}\;J_{4}} \big< J_{1}(1) \Big| V_{1} \Big| 
   J_{3}(1)\big> \nonumber 
\end{eqnarray}

\begin{eqnarray}
    \Big< \Big[J_{1} J_{2}\Big]^{J_{12}}\,J_{3},\, J \Big| \Big|
   &T_{12}^{(k)}  &
    \Big|\Big|\;  \Big[\bar{J}_{1} 
    \bar{J}_{2}\Big]^{\bar{J}_{12}}\,\bar{J_{3}},\,\bar{J}\Big> =  
      (-)^{2J_{12}+2\bar{J}+J_{3}+J_{12}+k+J}\;\delta_{J_{3}\bar{J}_{3}}
     \; \;\hat{J}\hat{\bar{J}}\nonumber\\&\times&W\big(J_{12}\,\bar{J_{12}}\,J\bar{J};k\,J_{3}\big)
      \;\; \big<J_{1} J_{2},\,{J_{12}} \Big|\Big| T_{12}^{(k)} 
    \Big|\Big| \bar{J}_{1} 
    \bar{J}_{2},\bar{J}_{12}\big>
\end{eqnarray}

\subsection{Surface Delta Interaction}

\begin{eqnarray}
<\ell \frac{1}{2} j\;||\;C_k \;||\;\ell^{'}  \frac{1}{2}  j'> &=& (-)^{\ell'+\ell}\sqrt{2j+1}\;\;C\begin{array}{ccc}j&k&j'\\ \frac{1}{2} & 0 & \frac{1}{2}\end{array}\;\;\;\;
\frac{1}{2}\left[1+(-)^{\ell+k+\ell'}\right]\label{eq:Ck}\\
<j_1 j_2 J T| \left(C_k\cdot C_k\right)    | j_3 j_4 J T >&=&(-)^{j_2+J+j_3}\left\{ \begin{array}{ccc}j_1&j_2&J\\j_4&j_3&k\end{array}\right\}<j_1\,||\;C_k\;||\,j_3> <j_2|| \;C_k \;||j_4> \nonumber\\
&=&(-)^{J+j_1+j_4}\; \hat{j_1}\,\hat{j_2}\;\left\{ \begin{array}{ccc}j_1&j_2&J\\j_4&j_3&k\end{array}\right\}
C\begin{array}{ccc}j_1&k&j_3\\\frac{1}{2}&0&\frac{1}{2}\end{array}\;\;C\begin{array}{ccc}j_2&k&j_4\\\frac{1}{2}&0&\frac{1}{2}\end{array}\hspace{2cm}
\end{eqnarray}
Alternately,

\begin{eqnarray}
<\ell \frac{1}{2} j||C_k ||\ell' \frac{1}{2}j'> &=& (-)^{\ell+\frac{1}{2}+j'+k}\;\;\hat{j}\hat{j^{'}}\left\{\begin{array}{ccc}\ell&\ell^{'}&k\\j'&j&\frac{1}{2}\end{array}\right\}<\ell || C_k || \ell'>\nonumber
\\
&=& (-)^{\ell+\frac{1}{2}+j'+k}\;\;\hat{j}\hat{j^{'}}
\left\{\begin{array}{ccc}\ell&\ell^{'}&k\\j'&j&\frac{1}{2}\end{array}\right\}
(-)^{\ell}\;\hat{\ell}\;\hat{\ell^{'}}
\left(\begin{array}{ccc}\ell&k&\ell^{'}\\0&0&0\end{array}\right)\label{eq:3j6j}
\\
&=& (-)^{\ell+\frac{1}{2}+j'+k}\;\;\hat{j}\hat{j^{'}}
\left\{\begin{array}{ccc}\ell&\ell^{'}&k\\j^{'}&j&\frac{1}{2}\end{array}\right\}
(-)^{\ell}\;\hat{\ell}\;\hat{\ell^{'}} (-)^{\ell-k}
C\begin{array}{ccc}\ell&k&\ell^{'}\\0&0&0\end{array}\nonumber
\\
&=& (-)^{3\ell+\frac{1}{2}+j^{'}}\;\;\hat{j}\;\hat{j^{'}}\;\hat{\ell}\;\hat{\ell^{'}}\;
\left\{\begin{array}{ccc}\ell&\ell^{'}&k\\j^{'}&j&\frac{1}{2}\end{array}\right\}
\; 
C\begin{array}{ccc}\ell&k&\ell^{'}\\0&0&0\end{array}\nonumber
\end{eqnarray}

Note that in Eq.(\ref{eq:3j6j}) one could make use of an identity
$$
\left(\begin{array}{ccc}j_1&j_2&J\\\frac{1}{2}&-\frac{1}{2}&0\end{array}\right) = -\hat{\ell_1}
\hat{\ell_2}\left(\begin{array}{ccc}\ell_1&\ell_2&J\\0&0&0\end{array}\right)
\left\{\begin{array}{ccc}j_1&j_2&J\\\ell_2&\ell_1&\frac{1}{2}\end{array}\right\}
$$
or
$$
(-)^{j_1+j_2+J}\left(\begin{array}{ccc}j_1&J&j_2\\\frac{1}{2}&0&-\frac{1}{2}\end{array}\right) = (-)^{1+\ell_1+\ell_2+J}\hat{\ell_1}
\hat{\ell_2}\left(\begin{array}{ccc}\ell_1&J&\ell_2\\0&0&0\end{array}\right)
\left\{\begin{array}{ccc}j_1&j_2&J\\\ell_2&\ell_1&\frac{1}{2}\end{array}\right\}
$$
or
$$
(-)^{j_1+j_2+J+1+\ell_2+J}\left(\begin{array}{ccc}j_1&J&j_2\\\frac{1}{2}&0&-\frac{1}{2}\end{array}\right) = (-)^{\ell_1}\hat{\ell_1}
\hat{\ell_2}\left(\begin{array}{ccc}\ell_1&J&\ell_2\\0&0&0\end{array}\right)
\left\{\begin{array}{ccc}j_1&j_2&J\\\ell_2&\ell_1&\frac{1}{2}\end{array}\right\}
$$
(with $\ell_1 + J + \ell_2$ even) and Eq.(\ref{eq:3j6j}) becomes

\begin{eqnarray}
(-)^{\ell+\frac{1}{2}+j'+k}\;\;\hat{j}\hat{j^{'}}
\left\{\begin{array}{ccc}\ell&\ell^{'}&k\\j'&j&\frac{1}{2}\end{array}\right\}
(-)^{\ell}\;\hat{\ell}\;\hat{\ell^{'}}
\left(\begin{array}{ccc}\ell&k&\ell^{'}\\0&0&0\end{array}\right)
&=&
(-)^{\ell+\frac{3}{2}+2j'+3k+j+\ell^{'}}\;\;\hat{j}\hat{j^{'}}\left(\begin{array}{ccc}j&k&j^{'}\\\frac{1}{2}&0&-\frac{1}{2}\end{array}\right)\nonumber
\\
&&\hspace{-2cm}=(-)^{\ell+\frac{3}{2}+2j'+3k+j+\ell^{'}}\;\;\hat{j}\hat{j^{'}}
\frac{1}{\hat{j'}}
(-)^{j-k+\frac{1}{2}}\;\;C\begin{array}{ccc}j&k&j^{'}\\\frac{1}{2}&0&\frac{1}{2}\end{array}\nonumber\\
&&\hspace{-2cm}=(-)^{\ell+1+\ell^{'}}\;\;\hat{j}
\;\;C\begin{array}{ccc}j&k&j^{'}\\\frac{1}{2}&0&\frac{1}{2}\end{array}\;.\nonumber
\end{eqnarray}
which agrees with Eq. (\ref{eq:Ck}) {\bf except for a minus...}

In LS Coupling
\begin{eqnarray}
<\ell_1\,\ell_2\, L \,S\, J \,| &\left(C_k\cdot C_k\right)& |\ell_3\,\ell_4 \,L\, S\,J\,> = \frac{1}{\hat{J}}<\,L S J \,|| \left(C_k\cdot C_k\right) ||\,L S J>\nonumber\\
&=& \hspace{-.5cm}(-)^{L+S+J} (2J+1) \frac{1}{\hat{J}}\left\{\begin{array}{ccc}j_1&j_2&J\\\ell_2&\ell_1&\frac{1}{2}\end{array}\right\} < \ell_1\,\ell_2 \,L\,|| \left(C_k\cdot C_k\right) ||\ell_3\,\ell_4 \,L\, >
\nonumber\\
&=& \hspace{-.5cm}(-)^{L+S+J}\;\hat{J}\; \left\{\begin{array}{ccc}j_1&j_2&J\\\ell_2&\ell_1&\frac{1}{2}\end{array}\right\}(-)^{\l_3+\ell_4}\;\hat{\ell_1}\hat{\ell_2}\hat{\ell_3}\hat{\ell_4}
\left(\begin{array}{ccc}\ell_1&k&\ell_3\\0&0&0\end{array}\right)
\left(\begin{array}{ccc}\ell_2&k&\ell_4\\0&0&0\end{array}\right)
\nonumber\\
&=& \hspace{-.5cm}(-)^{L+S+J+\ell_3+\ell_4+\ell_1+\ell_2+2k} 
\left\{\begin{array}{ccc}j_1&j_2&J\\\ell_2&\ell_1&\frac{1}{2}\end{array}\right\}\;\hat{J}\hat{\ell_1}\hat{\ell_2}\;
C\begin{array}{ccc}\ell_1&k&\ell_3\\0&0&0\end{array}\;
C\begin{array}{ccc}\ell_2&k&\ell_4\\0&0&0\end{array}
\nonumber
\end{eqnarray}

In all cases the off-diagonal matrix elements are related to the corresponding diagonal elements:  $M_{ij} = \sqrt{M_{ii}M_{jj}}\;$.

xxxxx

\subsection{SDI Matrix Elements - LS Coupling}

\subsubsection{Positive Parity, T=0}

\begin{eqnarray}
\,<L=0\;S=1\;J=1^+\;|V_{SDI}|\;LSJ=1> =\left( \begin{array}{ccc}s^2 & p^2&d^2 \\\hline
1& \sqrt{3}& \sqrt{5} \\
 \sqrt{3}& 3 & \sqrt{15}\\
\sqrt{5} &  \sqrt{15}& 5
\end{array}\right)
\end{eqnarray}

\subsubsection{Negative Parity, T=0}

\subsubsection{Positive Parity, T=1}

\paragraph{$\bf J=0^+$}

\begin{eqnarray}
\,<L=0\;S=0\;J=0^+\;|V_{SDI}|\;LSJ=0> =\left( \begin{array}{ccc}s^2 & p^2&d^2 \\\hline
1& \sqrt{3}& \sqrt{5} \\
 \sqrt{3}& 3 & \sqrt{15}\\
\sqrt{5} &  \sqrt{15}& 5
\end{array}\right)
\end{eqnarray}
$$Trace = 9$$

\paragraph{$\bf J=2^+$}

\begin{eqnarray}
\,<L=1\;S=1\;J=2^+\;|V_{SDI}|\;LSJ=2> =\left( \begin{array}{cc}p^2 & d^2\\\hline\\
\frac{2}{15}& 0 \\\\
0 & x \\
\end{array}\right)
\end{eqnarray}

\begin{eqnarray}
\,<L=0\;S=0\;J=2^+\;|V_{SDI}|\;LSJ=2> =\left( \begin{array}{ccc}p^2 & sd&d^2 \\\hline
\frac{2}{3}&  &  \\
  & \frac{19}{25} &  \\
\sqrt{\frac{12}{7}} &  -\sqrt{\frac{10}{7}}& \frac{818}{5^3 7}
\end{array}\right)
\end{eqnarray}

\begin{eqnarray}
\,<sd\;L=2\;S=1\;J=2^+\;|V_{SDI}|\;LSJ=2> =  \frac{6}{25}\\
\,< d^2\;L=2\;S=1\;J=2^+\;|V_{SDI}|\;LSJ=2> =
\end{eqnarray}
$$Trace = \frac{127}{35}$$

\paragraph{$\bf J=4^+$}
\begin{eqnarray}
\,<d^2\;L=3\;S=1\;J=4^+\;|V_{SDI}|\;LSJ=4> =\frac{4}{35}\\
< d^2\;L=4\;S=0\;J=4^+\;|V_{SDI}|\;LSJ=4> =\frac{26}{35}
\end{eqnarray}
$$Trace = \frac{6}{7}$$

\subsubsection{Negative Parity, T=1}

\begin{eqnarray}
\,<L=1\;S=1\;J=0^-\;|V_{SDI}|\;LSJ=0> =\left( \begin{array}{cc}ps & pd \\\hline
1& -\sqrt{2} \\
 -\sqrt{2}& 2 \\
\end{array}\right)
\end{eqnarray}

\begin{eqnarray}
\,<L=1\;S=1\;J=2^-\;|V_{SDI}|\;LSJ=2> =\left( \begin{array}{cc}ps & pd \\\hline
\frac{2}{5}& -\frac{\sqrt{8}}{5} \\
  -\frac{\sqrt{8}}{5}& \frac{4}{5} \\
\end{array}\right)
\end{eqnarray}

\begin{eqnarray}
\,<L=3\;S=1\;J=2^-\;|V_{SDI}|\;LSJ=2> =\frac{27}{35}
\end{eqnarray}

\begin{eqnarray}
\,<L=3\;S=1\;J=4^-\;|V_{SDI}|\;LSJ=4> =\frac{4}{7}
\end{eqnarray}

\subsection{SDI Matrix Elements - jj Coupling}

\subsection{Positive Parity, T=1}

\begin{eqnarray}
\,<j_1 j_2;J=0^+\;T=1\;|V_{SDI}|\;j_3j_4; J=0^+ T=1> =\left( \begin{array}{ccccc}s_{\frac{1}{2}}s_{\frac{1}{2}} &p_{\frac{1}{2}}p_{\frac{1}{2}}& p_{\frac{3}{2}}p_{\frac{3}{2}} & d_{\frac{3}{2}}d_{\frac{3}{2}} & d_{\frac{5}{2}}d_{\frac{5}{2}} 
\\\hline\\
1& -1& -\sqrt{2} & \sqrt{2} &    \sqrt{3}\\&&&&\\
-1  & 1 & \sqrt{2} &  -\sqrt{2}  & -\sqrt{3}\\\\
  -\sqrt{2}   & \sqrt{2}  &2&  -2&  -\sqrt{6}\\\\
 \sqrt{2} &  -\sqrt{2} &  -2 & 2& \sqrt{6} \\\\
\sqrt{3}  & -\sqrt{3} & -\sqrt{3} & \sqrt{6} &3
\end{array}\right)
\end{eqnarray}

\paragraph{ $\bf j_1 j_2;J=2^+\;$}

\begin{eqnarray}
\,<j_1 j_2;J=2^+\;|V_{SDI}|\;j_3j_4; J=2^+> =\left( \begin{array}{ccccccc}p_{\frac{1}{2}}p_{\frac{3}{2}} &p_{\frac{3}{2}}p_{\frac{3}{2}}& s_{\frac{1}{2}}d_{\frac{3}{2}} & d_{\frac{3}{2}}d_{\frac{3}{2}} & s_{\frac{1}{2}}d_{\frac{5}{2}}& d_{\frac{3}{2}}d_{\frac{5}{2}}& d_{\frac{5}{2}}d_{\frac{5}{2}}
\\\hline\\
\frac{2}{5}& -\frac{2}{5}&-\frac{2}{5}& \frac{2}{5} &-\frac{\sqrt{6}}{5} & -\frac{2\sqrt{21}}{35}& \frac{4\sqrt{21}}{35}
\\&&&&\\
-\frac{2}{5} & \frac{2}{5} & \frac{2}{5} & -\frac{2}{5}& \frac{\sqrt{6}}{5} & \frac{2\sqrt{21}}{35}& -\frac{4\sqrt{21}}{35}
\\\\
 -\frac{2}{5}  & \frac{2}{5}  & \frac{2}{5}&  -\frac{2}{5} & \frac{\sqrt{6}}{5} & \frac{2\sqrt{21}}{35} &  -\frac{4\sqrt{21}}{35}
\\\\
 \frac{2}{5}&  -\frac{2}{5}&  -\frac{2}{5}& \frac{2}{5} &  -\frac{\sqrt{6}}{5}& -\frac{2\sqrt{21}}{35}& \frac{4\sqrt{21}}{35} 
\\\\
-\frac{\sqrt{6}}{5} & \frac{\sqrt{6}}{5} &  \frac{\sqrt{6}}{5} &  -\frac{\sqrt{6}}{5} & \frac{3}{5} & \frac{3\sqrt{14}}{35} & -\frac{6\sqrt{14}}{35} 
\\\\
-\frac{2\sqrt{21}}{35} & \frac{2\sqrt{21}}{35} & \frac{2\sqrt{21}}{35} & -\frac{2\sqrt{21}}{35} & \frac{3\sqrt{14}}{35} & \frac{6}{35} &  -\frac{12}{35} 
\\\\
\frac{4\sqrt{21}}{35} & -\frac{4\sqrt{21}}{35} & -\frac{4\sqrt{21}}{35} & \frac{4\sqrt{21}}{35} & -\frac{6\sqrt{14}}{35} &  -\frac{12}{35} & \frac{24}{35}
\end{array}\right)
\end{eqnarray}
$$Trace = \frac{107}{35}$$

\paragraph{ $\bf j_1 j_2;J=4^+\;$}

\begin{eqnarray}
\,<j_1 j_2;J=4^+\;T=1\;|V_{SDI}|\;j_3j_4; J=4^+ T=1> =\left( \begin{array}{cc} d_{\frac{3}{2}}d_{\frac{5}{2}} & d_{\frac{5}{2}}d_{\frac{5}{2}} 
\\\hline\\
\frac{4}{7} & -\frac{\sqrt{8}}{7}\\\\
 -\frac{\sqrt{8}}{7}  & \frac{2}{7} \\\\
\end{array}\right)
\end{eqnarray}
$$Trace = \frac{6}{7}$$

Negative Parity:

\begin{eqnarray}
\,<j_1 j_2;J=1^-\;|V_{SDI}|\;j_3j_4; J=1^-> =\left( \begin{array}{ccccc}s_{\frac{1}{2}}p_{\frac{1}{2}} &s_{\frac{1}{2}}p_{\frac{3}{2}}& p_{\frac{1}{2}}d_{\frac{3}{2}} & p_{\frac{3}{2}}d_{\frac{3}{2}} & p_{\frac{3}{2}}d_{\frac{5}{2}} 
\\\hline\\
\frac{1}{3}& \frac{\sqrt{2}}{5}& -\frac{\sqrt{2}}{3} & -\frac{\sqrt{10}}{15} &    -\sqrt{\frac{2}{5}}\\&&&&\\
\frac{\sqrt{2}}{5}  & \frac{2}{3} &  -\frac{2}{3} & -\frac{2\sqrt{5}}{15}  & -\sqrt{\frac{4}{5}}\\\\
  -\frac{\sqrt{2}}{3}   & -\frac{2}{3}  & \frac{2}{3}&  \frac{2\sqrt{5}}{15}&  \sqrt{\frac{4}{5}}\\\\
 -\frac{\sqrt{10}}{15}& -\frac{2\sqrt{5}}{15} &  \frac{2\sqrt{5}}{15}& \frac{2}{15}& \frac{2}{5} \\\\
-\sqrt{\frac{2}{5}}\ & -\sqrt{\frac{4}{5}} & \sqrt{\frac{4}{5}} & \frac{2}{5} & \frac{6}{5}
\end{array}\right)
\end{eqnarray}

\begin{eqnarray}
\,<j_1 j_2;J=3^-\;|V_{SDI}|\;j_3j_4; J=3^-> =\left( \begin{array}{ccc}p_{\frac{3}{2}}d_{\frac{3}{2}}& p_{\frac{1}{2}}d_{\frac{5}{2}}  & p_{\frac{3}{2}}d_{\frac{5}{2}} 
\\\hline\\
\frac{18}{35}& -\frac{3}{7}\sqrt{\frac{6}{5}}& \frac{6\sqrt{6}}{35} \\&&\\
-\frac{3}{7}\sqrt{\frac{6}{5}}  & \frac{3}{7} &  -\frac{6\sqrt{5}}{35} \\\\
 \frac{6\sqrt{6}}{35}   &  -\frac{6\sqrt{5}}{35} & \frac{12}{35}\\\\
 
\end{array}\right)
\end{eqnarray}

\newpage
%If $k=0$,

\begin{table}[h]
\centering

\begin{tabular}{|c||c|c|}\hline
index&$[f]_{cs}$&$(\lambda \mu)_{cs} S$\\\hline
1&[222]&(0 0) 1\\
2&[222]&(0 0) 3\\
3&[2211]&(0 0) 0\\
4&[2211]&(0 0) 2\\\hline
5&[221]&(0 1)$\frac{1}{2}$\\
6&[221]&(0 1)$\frac{3}{2}$\\
7&[221]&(0 1)$\frac{5}{2}$\\
8&[2111]&(0 1)$\frac{1}{2}$\\
9&[2111]&(0 1)$\frac{3}{2}$\\\hline
10&[22]&(0 2) 0\\
11&[22]&(0 2) 2\\
12&[22]&(1 0) 1\\
13&[211]&(0 2) 1\\
14&[211]&(1 0) 0\\
15&[211]&(1 0) 1\\
16&[211]&(1 0) 2\\
17&[1111]&(0 2) 0\\
18&[1111]&(1 0) 1\\\hline
19&[21]&(1 1)$\frac{1}{2}$\\
20&[21]&(1 1)$\frac{3}{2}$\\
21&[21]&(0 0)$\frac{1}{2}$\\
22&[111]&(1 1)$\frac{1}{2}$\\
23&[111]&(0 0)$\frac{3}{2}$\\\hline
24&[2]&(2 0) 1\\
25&[2]&(0 1) 0\\
26&[11]&(2 0) 0\\
27&[11]&(0 1) 1\\\hline
28&[1]&(1 0) 0\\
29&[]&(0 0) 1\\
\hline\end{tabular}

\caption{Index of $SU(6)_{cs}\; (\lambda \mu)_{c}\; S$ representations.}  
\label{tab:index}
\end{table}

\begin{table}[h]
\centering

\begin{tabular}{|c|c|c|c|c|}\hline
$ [22]    \; ( 0 2)_c\;  2S= 0\;\rightarrow [21]   $ &    & 1 &    &    \\
$ [22]    \; ( 0 2)_c\;  2S= 4\;\rightarrow [21]   $ &    &    & 1 &    \\
$ [22]    \; ( 1 0)_c\;  2S= 2\;\rightarrow [21]   $ &    & 0.57735027 &-0.57735027 &-0.57735027 \\
     \hline
$ [211]   \; ( 0 2)_c\;  2S= 2\;\rightarrow [21]   $ &    & 0.70710678 &-0.70710678 &    \\
$ [211]   \; ( 1 0)_c\;  2S= 0\;\rightarrow [21]   $ &    & 0.70710678 &    &-0.70710678 \\
$ [211]   \; ( 1 0)_c\;  2S= 2\;\rightarrow [21]   $ &    & 0.81649658 & 0.40824829 & 0.40824829 \\
$ [211]   \; ( 1 0)_c\;  2S= 4\;\rightarrow [21]   $ &    &    & 1 &    \\
\hline\end{tabular}
\caption{Cfps for four particles $SU(6)_{cs}\; (\lambda \mu)_{c}\; S$ 
representations $\rightarrow [21]$.}  
\label{tab:cfps4-21}
\end{table}
\begin{table}[h]
\centering

\begin{tabular}{|c|c|c|}\hline

$ [211]   \; ( 0 2)_c\;  2S= 2\;\rightarrow [111]  $ &-1 &    \\
$ [211]   \; ( 1 0)_c\;  2S= 0\;\rightarrow [111]  $ &-1 &    \\
$ [211]   \; ( 1 0)_c\;  2S= 2\;\rightarrow [111]  $ & 0.57735027 &-0.81649658 \\
$ [211]   \; ( 1 0)_c\;  2S= 4\;\rightarrow [111]  $ &    & 1 \\
\hline\end{tabular}
\caption{Cfps for four particles $SU(6)_{cs}\; (\lambda \mu)_{c}\; S$ 
representations $\rightarrow [111]$.}  
\label{tab:cfps4-111}
\end{table}

\begin{table}[h]
\centering

\begin{tabular}{|c||c|c|c|}\hline
$ [221]   \; ( 0 1)_c\;  2S= 1\;\rightarrow [22]   $  & 0.86602540 &  &-0.5 \\
$ [221]   \; ( 0 1)_c\;  2S= 3\;\rightarrow [22]   $  &  & 0.61237244 & 0.79056942 \\
$ [221]   \; ( 0 1)_c\;  2S= 5\;\rightarrow [22]   $  &  & 1 &  \\
\hline\end{tabular}
\caption{Cfps for  five particles $SU(6)_{cs}\; (\lambda \mu)_{c}\; 
S$ representations $\rightarrow [22]$.}  
\label{tab:cfps5a}
\end{table}

\begin{table}[h]
\centering

\begin{tabular}{|c||c|c|c|c|}\hline
 $ [221]  \; ( 0 1)_c\;  2S= 1\;\rightarrow [211]  $  & 0.40824829 &-0.40824829 &-0.81649658 &  \\
$ [221]   \; ( 0 1)_c\;  2S= 3\;\rightarrow [211]  $  &-0.64549722 &  &-0.64549722 & 0.40824829 \\
$ [221]   \; ( 0 1)_c\;  2S= 5\;\rightarrow [211]  $ &   & 
 &  &-1 \\\hline
$ [2111]  \; ( 0 1)_c\;  2S= 1\;\rightarrow [211]  $ & 0.57735027 &-0.57735027 & 0.57735027 &  \\
$ [2111]  \; ( 0 1)_c\;  2S= 3\;\rightarrow [211]  $ &  0.73029674 &  &-0.36514837 & 0.57735027 \\
 \hline\end{tabular}
\caption{Cfps for  five particles $SU(6)_{cs}\; (\lambda \mu)_{c}\; 
S$ representations $\rightarrow [211]$.}  
\label{tab:cfps5b}
\end{table}

\begin{table}[h]
\centering

\begin{tabular}{|c||c|c|}\hline
$ [2111]  \; ( 0 1)_c\;  2S= 1\;\rightarrow [1111] $ &-0.77459667 &-0.63245553 \\
$ [2111]  \; ( 0 1)_c\;  2S= 3\;\rightarrow [1111] $ &   & 1  \\
 \hline\end{tabular}
\caption{Cfps for  five particles $SU(6)_{cs}\; (\lambda \mu)_{c}\; S$ representations
 $\rightarrow [1111]$.}  
\label{tab:cfps5c}
\end{table}

\begin{table}[h]
\centering

\begin{tabular}{|c||c|c|c|}\hline
  &()&()&( )\\\hline
$ [222]   \; ( 0 0)_c\;  2S= 2\;\rightarrow [221]  $ & 0.74535599 & 0.66666667 &     \\
$ [222]   \; ( 0 0)_c\;  2S= 6\;\rightarrow [221]  $ &            
&            & 1 \\\hline
$ [2211]  \; ( 0 0)_c\;  2S= 0\;\rightarrow [221]  $ & 1  &            &            \\
$ [2211]  \; ( 0 0)_c\;  2S= 4\;\rightarrow [221]  $ &            & 0.8 &-0.6 \\
\hline\end{tabular}
\caption{Cfps for  six particles $SU(6)_{cs}\; (\lambda \mu)_{c}\; S$ representations
 $\rightarrow [2211]$.}  
\label{tab:cfps6a}
\end{table}

\begin{table}[h]
\centering

\begin{tabular}{|c||c|c|}\hline
  &[11]&[11]\\\hline
$ [2211]  \; ( 0 0)_c\;  2S= 0\;\rightarrow [2111] $ &-1  &   \\
$ [2211]  \; ( 0 0)_c\;  2S= 4\;\rightarrow [2111] $ &    & 1 \\
\hline\end{tabular}
\caption{Cfps for  six particles $SU(6)_{cs}\; (\lambda \mu)_{c}\; S 
$ representations $\rightarrow [2111]$.}  
\label{tab:cfps6b}
\end{table}

\newpage
\include{tables}

\end{document}